\documentclass[usenatbib,letterpaper]{mn2e}
\usepackage[totalwidth=480pt,totalheight=680pt,twoside=false,voffset=.5cm]{geometry}

\usepackage{graphics,graphicx,epsfig,ulem,multirow,appendix,lscape,amsmath,amssymb,color,enumitem,natbib,url}

\usepackage{upgreek}
\usepackage[none]{hyphenat}	

\long\def\symbolfootnote[#1]#2{\begingroup%
\def\thefootnote{\fnsymbol{footnote}}\footnote[#1]{#2}\endgroup} 

\urldef{\sofiaFIFIHandbook}\url{https://www.sofia.usra.edu/science/proposing-and-observing/observers-handbook-cycle-8/3-fifi-ls/31-specifications#Table%20of%20ContentsFIFI-LSspecifications}

\begin{document}

\title[Hot dust trapped in gas around main-sequence stars]
  {Gas trapping of hot dust around main-sequence stars}
\author[T. D. Pearce, A. V. Krivov and M. Booth]
  {Tim D. Pearce\thanks{timothy.pearce@uni-jena.de}, Alexander V. Krivov and Mark Booth\\
  Astrophysikalisches Institut und Universit\"{a}tssternwarte, Friedrich-Schiller-Universit\"{a}t Jena, Schillerg\"{a}{\ss}chen 2-3, D-07745 Jena, Germany}
\date{Released 2002 Xxxxx XX}
\pagerange{\pageref{firstpage}--\pageref{lastpage}} \pubyear{2002}

\def\LaTeX{L\kern-.36em\raise.3ex\hbox{a}\kern-.15em
    T\kern-.1667em\lower.7ex\hbox{E}\kern-.125emX}

\newtheorem{theorem}{Theorem}[section]

\label{firstpage}

\maketitle


\begin{abstract}              
In 2006 Vega was discovered to display excess near-infrared emission. Surveys now detect this phenomenon for one fifth of main-sequence stars, across various spectral types and ages. The excesses are interpreted as populations of small, hot dust grains very close to their stars, which must originate from comets or asteroids. However, the presence of such grains in copious amounts is mysterious, since they should rapidly sublimate or be blown out of the system. Here we investigate a potential mechanism to generate excesses: dust migrating inwards under radiation forces sublimates near the star, releasing modest quantities of gas which then traps subsequent grains. This mechanism requires neither specialised system architectures nor high dust supply rates, and could operate across diverse stellar types and ages. The model naturally reproduces many features of inferred dust populations, in particular their location, preference for small grains, steep size distribution, and dust location scaling with stellar luminosity. For Sun-like stars the mechanism can produce ${2.2 \; \mu {\rm m}}$ excesses that are an order of magnitude larger than those at ${8.5 \; \mu {\rm m}}$, as required by observations. However, for A-type stars the simulated near-infrared excesses were only twice those in the mid infrared; grains would have to be ${5-10}$ times smaller than those trapped in our model to be able to explain observed near-infrared excesses around A stars. Further progress with any hot dust explanation for A stars requires a means for grains to become very hot without either rapidly sublimating or being blown out of the system.

\end{abstract}

\begin{keywords}
planetary systems, planetary systems: zodiacal dust, stars: circumstellar matter
\end{keywords}


\section{Introduction}
\label{sec: Introduction}

\noindent In the last 15 years excess near-infrared (NIR) emission has been detected around 20 per cent of main-sequence stars, across AFGK spectral types and diverse stellar ages \citep{Absil2006, Absil2013, Ertel2014, Mennesson2014, Ertel2016, Nunez2017, Ertel2018, Ertel2020}. Unlike the ubiquitous NIR excesses around young, pre-main-sequence stars, which stem from hot dust immersed in planet-forming discs, the origin of NIR excesses around older main-sequence stars is unknown. These excesses are attributed to dust at temperatures of order ${1000 \: {\rm K}}$, and thus located very close to the stars. This dust could originate from cometary or asteroid belts located farther out in the planetary systems, yet there are no clear correlations between NIR excesses and those at longer wavelengths that would be associated with large belts \citep{MillanGabet2011, Ertel2014, Mennesson2014, Ertel2018, Ertel2020}. Spectral energy distribution (SED) modelling and polarization measurements infer the dust to have a steep size distribution with an overabundance of small grains, and be located near the sublimation region of the star \citep{diFolco2007, Akeson2009, Defrere2011, Defrere2012, Lebreton2013, Marshall2016, Kirchschlager2017}. The dust grains are likely carbonaceous \citep{Absil2006, Sezestre2019}, and there is a significant trend for dust location to increase with stellar luminosity \citep{Kirchschlager2017}.

However, the presence of such grains in copious amounts is difficult to explain, since they should rapidly sublimate or blow out of the system. Many models have been proposed, but none have yet explained both the phenomenon and its ubiquity. The dust cannot be produced in a conventional steady-state collisional cascade close to the star, due to the short timescales at these distances \citep{Wyatt2007, Lebreton2013}. Comets could deposit material close to the star, but such models struggle to produce the required low mid-infrared (MIR) excesses and steep grain size distributions \citep{Sezestre2019}, and may require specific system architectures \citep{Bonsor2014, Raymond2014, Faramaz2017}. Transient dynamical upheavals, such as the Late Heavy Bombardment, could supply material to the very inner regions of the system; however, such events are too short-lived to explain the frequency of observations \citep{Bonsor2013}. Material migrating inwards through Poynting-Robertson (PR) drag and piling up at the sublimation zone cannot produce enough hot dust to reproduce observations, since the time that small grains remain near the sublimation region is too short \citep{vanLieshout2014, Sezestre2019}.  Trapping grains via the Differential Doppler Effect is ineffective against PR-drag \citep{Sezestre2019}, and whilst magnetic trapping has been proposed \citep{Rieke2016} the effectiveness of the mechanism is unclear \citep{Kral2017Review, Kimura2020}.

In this paper we investigate a potential mechanism to generate NIR excesses across diverse stellar types and ages, without requiring specific system architectures or exceeding constraints on MIR emission: dust migrating inwards under radiation forces sublimates near the star, releasing gas which then traps subsequent grains. These hot grains, the source of the near-infrared emission, are held on stable orbits by gas. The idea was briefly explored before \citep{Lebreton2013}, but the required gas quantity appeared too high to be compatible with observations \citep{Su2016}. We perform detailed analyses and show that the gas quantities needed are much smaller than the previous estimate.

The work is laid out as follows. In Section \ref{sec: dustEvolutionNoGas} we describe the evolution of dust grains close to the star in the absence of gas. In Section \ref{sec: dustEvolutionGas} we outline our gas model, and show that grains can become trapped in a gas disc. We examine the implications of our simulations, including observational comparisons, in Section \ref{sec: dustLifetimeAndObservationalComparison}. We discuss the results in Section \ref{sec: discussion}, and conclude in Section \ref{sec: conclusions}.


\section{Hot dust evolution in the absence of a gas disc}
\label{sec: dustEvolutionNoGas}

\noindent We first describe the evolution of hot dust close to the star in the absence of a gas disc, under the influence of radiation forces and sublimation. This evolution cannot keep hot dust close to the star for long enough to be compatible with NIR observations \citep{vanLieshout2014, Sezestre2019}, but this evolution must be considered to put the gas model in context.

\subsection{Grain evolution model without a gas disc}
\label{subsec: modellingNoGas}

\noindent Here we describe our model for dust grain evolution without gas. We model a single grain in orbit around a star, subject to gravity, radiation forces and sublimation. We assume the grain to originate somewhere exterior to the sublimation region, but make no assumptions about the exact source (e.g. a collisional debris disc or cometary disintegration). We then follow the evolving size and location of the grain.

\subsubsection{Forces}
\label{subsec: forces}

\noindent In the absence of gas, the dust grain is subject to gravity and radiation forces. Its equation of motion is

\begin{equation}
{\bf F} = {\bf F}_{\rm grav} + {\bf F}_{\rm rad},
\label{eq: equationOfMotionNoGas}
\end{equation}

\noindent where ${\bf F}$ is the total force and ${\bf F}_{\rm grav}$ and ${\bf F}_{\rm rad}$ are those due to gravity and radiation, respectively. The gravitational force is

\begin{equation}
{\bf F}_{\rm grav} = -\frac{G M_\star m}{r^2} {\bf \hat{r}},
\label{eq: gravitationalForce}
\end{equation}

\noindent where $G$, $M_*$, $m$, $r$ and ${\bf \hat{r}}$ are the gravitational constant, star mass, dust grain mass, star-grain separation and star-grain separation unit vector, respectively.

The combined effects of radiation pressure and PR-drag induce a force ${\bf F}_{\rm rad}$ on the grain:

\begin{equation}
{\bf F}_{\rm rad} = \beta |{\bf F}_{\rm grav}| \left[\left(1 - \frac{\dot{r}_{\rm d}}{c} \right){\bf \hat{r}} - \frac{{\bf v}_{\rm d}}{c}\right],
\label{eq: radiationForce}
\end{equation}

\noindent where $\beta$ is the ratio of radiation pressure to the gravitational force, ${\bf v}_{\rm d}$ is the dust velocity, $\dot{r}_{\rm d}$ is the radial component of ${\bf v}_{\rm d}$, $\bf \hat{r}$ is the radial unit vector and $c$ is the speed of light \citep{Burns1979}. Radiation pressure counteracts gravity, so a grain on a circular orbit has a reduced velocity 

\begin{equation}
\mathbf{v}_{\rm d} = {\bf v}_{\rm Kep} \sqrt{1 - \beta},
\label{eq: dustVelocity}
\end{equation}

\noindent where ${\bf v}_{\rm Kep}$ is the Keplerian orbital velocity.

\subsubsection{Sublimation}
\label{subsec: sublimation}

\noindent We implement the sublimation prescription of \citet{Lebreton2013}, based on that of \citet{Lamy1974}. Particles evaporate from a grain at temperature $T_{\rm d}$ at a rate

\begin{equation}
f_{\rm evap} = \frac{P_{\rm eq}}{\sqrt{2 \pi \mu m_{\rm u} k_{\rm B} T_{\rm d}}},
\label{eq: evaporationFlux}
\end{equation}

\noindent where $f_{\rm evap}$ is the flux of particles evaporating from the grain surface, $P_{\rm eq}$ and $\mu$ are the saturation pressure and molecular weight of evaporating material respectively, $m_{\rm u}$ is the atomic mass unit and $k_{\rm B}$ is the Boltzmann constant. In the absence of a gas disc, the grain mass thus evolves at a rate

\begin{equation}
\frac{{\rm d}m}{{\rm d}t} = -\gamma 4 \pi s^2 f_{\rm evap} \mu m_{\rm u},
\label{eq: grainMassEvaporation}
\end{equation}

\noindent where $s$ is the grain radius and $\gamma$ a laboratory-derived factor to account for the processes being less than 100 per cent efficient; we use $\gamma = 0.7$ as in \citet{Sezestre2019}, noting that $\gamma$ is denoted $\alpha$ in that paper. 

The ideal gas law relates pressure $P$ to density $\rho$ and temperature $T$ by ${P = \rho k_{\rm B} T / (\mu m_{\rm u})}$. Substituting ${{\rm d}m = 4 \pi s^2 \rho_{\rm d} {\rm d}s}$, where $\rho_{\rm d}$ is the dust grain density, yields the sublimation rate as

\begin{equation}
\frac{{\rm d}s}{{\rm d}t} = - \gamma \sqrt{\frac{k_{\rm B} T_{\rm d}}{2 \pi \mu m_{\rm u}}} \frac{\rho_{\rm eq}}{\rho_{\rm d}}.
\label{eq: sublimationRateNoGas}
\end{equation}

\noindent This is equivalent to Equation 17 of \citet{Lebreton2013} if no gas disc is present. The density of evaporating gas at saturation pressure, $\rho_{\rm eq}$, is found from the Clausius-Clapeyron equation

\begin{equation}
\log_{10}\left( \frac{\rho_{\rm eq}}{\rm g \: cm^{-3}}\right)= B - A\left( \frac{T_{\rm d}}{\rm K}\right)^{-1} - \log_{10}\left(\frac{T_{\rm d}}{\rm K}\right),
\label{eq: gasDensityAtSaturationPressure}
\end{equation}

\noindent where $A$ and $B$ are material-specific quantities determined empirically \citep{Lebreton2013}; for carbon, $A = 37215$ and $B = 7.2294$ \citep{Zavitsanos1973, Sezestre2019}. Hence $\rho_{\rm eq}$ is extremely sensitive to $T_{\rm d}$.

\subsubsection{Dust grain properties}
\label{subsec: dustGrainProperties}

\noindent SED modelling finds that pure silicate grains fail to reproduce hot excess observations, whilst carbonaceous materials are compatible with the data \citep{Absil2006, Akeson2009, Kirchschlager2017}. For simplicity we therefore model dust as solid, non-porous spheres of pure carbon, with density ${2 \: {\rm g \: cm}^{-3}}$. Gas that sublimates off the dust is assumed to be atomic carbon, with molecular weight 12.01.

The dust $\beta$ and temperature are calculated from Equations 3 and 14 in \citet{Gustafson1994}, respectively; since we assume homogeneous spheres, we use Mie theory to calculate the radiation pressure and absorption efficiencies. We use optical properties for ${1000\degr{\rm C}}$ carbon as measured by \citet{Jager1998}\footnote{\url{https://www.astro.uni-jena.de/Laboratory/OCDB/carbon.html}}. A grid of dust temperature and $\beta$ were pre-calculated for different dust sizes and distances, and then interpolated in calculations.

\subsubsection{Stellar properties}
\label{subsec: stellarProperties}

\noindent We examine the interaction around Sun-like and Vega-like stars (i.e. stars of spectral type G2 and A0, respectively). We use stellar properties from Mamajek's online tables\footnote{\url{http://www.pas.rochester.edu/~emamajek/EEM_dwarf_UBVIJHK_colors_Teff.txt}} \citep{Pecaut2013}; the G2 star has mass ${1.02 \: {\rm M}_\odot}$, radius ${1.01 \: {\rm R}_\odot}$ and bolometric luminosity ${1.02 \: {\rm L}_\odot}$, and the corresponding A0 star values are ${2.30 \: {\rm M}_\odot}$, ${2.09 \: {\rm R}_\odot}$ and ${34.7 \: {\rm L}_\odot}$, respectively. Stellar spectra were taken from \citet{Kurucz1992}.

\subsection{Numerical implementation}
\label{subsec: numericalImplementation}

\noindent We implement the above physics using a bespoke two-dimensional, fourth-order Runge-Kutta integrator with variable step size. This models the evolution of a single dust grain around a star, until the grain either completely sublimates, collides with the star, or reaches a distance of ${100 \: {\rm au}}$ (i.e. is expelled from the system). The variable time step is 0.01 times the smaller of the instantaneous orbital and sublimation timescales, where the scale factor 0.01 was found to be sufficiently small for convergence. The program well-reproduces literature results for grain evolution in the absence of gas, as described in Section \ref{subsec: resultsNoGas}.


\subsection{Results in the absence of gas}
\label{subsec: resultsNoGas}

\begin{figure}
  \centering
   \includegraphics[width=7cm]{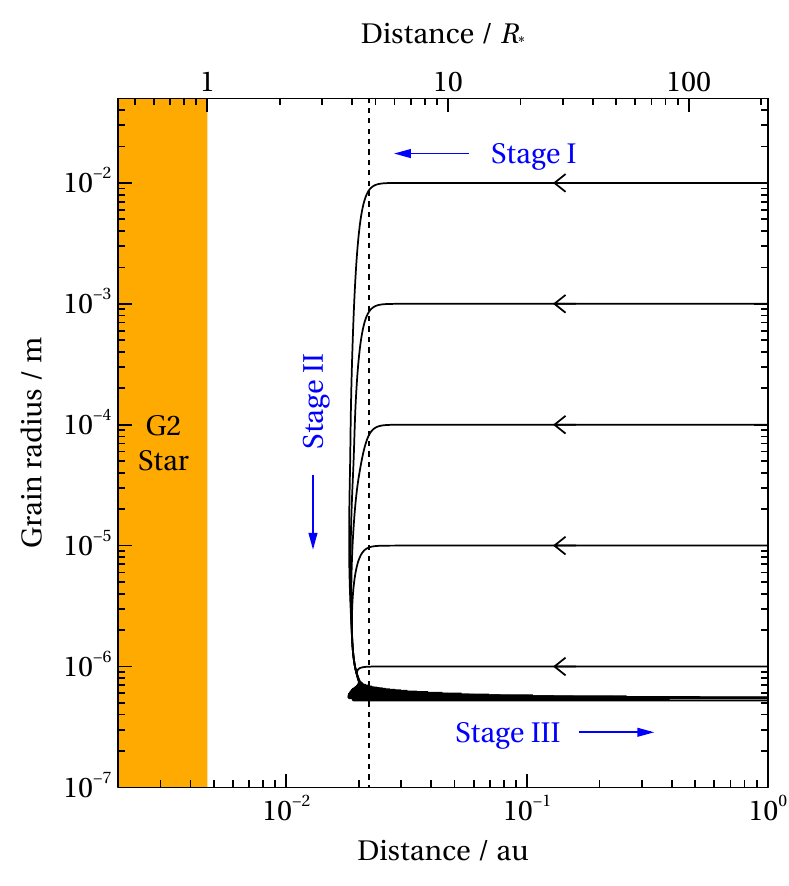}
   \caption{Evolution of solid carbon grains of different initial sizes orbiting a G2 star, if no gas is present. The qualitative evolution is independent of initial grain size. First, grains migrate inwards under PR-drag, maintaining circular orbits (Stage I). When grains approach the star they begin to sublimate over a narrow distance range (Stage II). As grain radii decrease their $\beta$ values increase, and radiation forces cause their orbital eccentricities to grow (Stage III). Eventually grains leave the system. The Figure is very similar to Figures 4 and 5 in \citet{Sezestre2019}. The dashed line is at ${r_{\rm s} = 0.022 \: {\rm au}}$, where grains of all initial sizes have lost 50 per cent of their mass through sublimation; this is taken as the gas input location in later modelling. Note that dust arbitrarily originates at ${1 \: {\rm au}}$ on this figure, but provided it starts exterior to the sublimation region the fundamental results are unchanged.}
   \label{fig: grainRadiusVsDistanceNoGas}
\end{figure}

\noindent We now describe the evolution of a dust grain under PR-drag and sublimation, in the absence of gas. The behaviour is qualitatively identical for G2 and A0 stars and initial grain sizes, with an example shown on Figure \ref{fig: grainRadiusVsDistanceNoGas}. The evolution follows a 3-stage process, as repeatedly demonstrated in the literature (e.g. \citealt{Krivov1998, Kobayashi2008, Sezestre2019}). First, the grain migrates towards the star under PR-drag (Stage I on Figure \ref{fig: grainRadiusVsDistanceNoGas}). As the grain approaches the sublimation region, its size is rapidly reduced over a very small distance range (Stage II). Finally, the grain eccentricity grows as reducing its size has increased its $\beta$ (Stage III). PR-drag continues to pull the grain pericentre closer to the star and the grain continues to sublimate, further increasing its eccentricity. Eventually the grain becomes so small that it becomes unbound and leaves the system; in our simulations its radius at blowout is ${0.5 \: \mu {\rm m}}$ for a G2 star, and ${5 \: \mu {\rm m}}$ for an A0 star.

Equations \ref{eq: sublimationRateNoGas} and \ref{eq: gasDensityAtSaturationPressure} show that, in the absence of gas, some amount of sublimation occurs at all distances. However, the majority of sublimation occurs over a very narrow distance range (Figure \ref{fig: grainRadiusVsDistanceNoGas}). We therefore consider a characteristic sublimation distance $r_{\rm s}$, defined as the distance at which a grain has lost 50 per cent of its mass (i.e. the radius of a spherical grain is $0.79$ times its original radius). We find that for solid carbon grains orbiting a given star type, $r_{\rm s}$ is roughly constant for all initial grain sizes. For a G2 star ${r_{\rm s} = 0.022 \: {\rm au}}$, and for an A0 star ${r_{\rm s} = 0.13 \: {\rm au}}$. These characteristic sublimation radii will be used when considering dust evolution in the presence of gas.


\section{Hot dust evolution in the presence of a gas disc}
\label{sec: dustEvolutionGas}

\noindent We now build on the previous section by considering grain evolution if a gas disc is also present. The idea is that grains migrate inwards through PR-drag and sublimate as before, but the sublimated gas now affects subsequent grains. 

\subsection{Grain evolution model with gas}
\label{subsec: modellingGas}

Here we describe our model for dust evolution in the presence of a gas disc.


\subsubsection{Gas disc properties}
\label{subsec: gasDiscProperties}

\noindent We assume that a gas disc is present around the sublimation radius, which is produced and maintained by dust sublimation. Although sublimation occurs over a very narrow distance range (Figure \ref{fig: grainRadiusVsDistanceNoGas}), gas is expected to move away from its creation point and form a disc rather than a narrow ring. Here we describe the assumed gas disc; some properties may be simplifications, but should suffice to demonstrate the viability of the model.

We assume the disc to be much less massive than the star, but sufficiently dense for gas to behave viscously. We also impose that the disc is vertically thin and isothermal. Using these assumptions we can describe the gas with a thin protoplanetary disc model summarised in \citet{Armitage2010}.

The gas undergoes bulk circular motion about the star, where pressure-support causes the bulk gas velocity ${\bf v}_{\rm gas}$ to be slightly smaller than the Keplerian orbital velocity ${\bf v}_{\rm Kep}$:

\begin{equation}
{\bf v}_{\rm gas} = {\bf v}_{\rm Kep} \sqrt{1-\eta},
\label{eq: gasVelocity}
\end{equation}

\noindent where

\begin{equation}
\eta \equiv \left(\frac{c_{\rm s}}{v_{\rm Kep}}\right)^2
\label{eq: eta}
\end{equation}

\noindent and $c_{\rm s}$ is the isothermal sound speed, given by

\begin{equation}
c_{\rm s} = \sqrt{\frac{k_{\rm B} T_{\rm gas}}{\mu m_{\rm u}}}.
\label{eq: soundSpeed}
\end{equation}

\noindent Since gas is assumed to be produced by dust sublimation, its molecular weight $\mu$ is assumed to be that of the dust. For the parameters in our model the factor ${\eta \ll 1}$, so the bulk gas velocity is approximately Keplerian.

For a thin disc the midplane gas density $\rho_{\rm gas}$ is related to surface density $\Sigma$ by

\begin{equation}
\rho_{\rm gas} = \frac{\Sigma}{\sqrt{2 \pi} H}.
\label{eq: densityToSurfaceDensityRelation}
\end{equation}

\noindent The scale height $H$ is

\begin{equation}
H = \frac{c_{\rm s}}{\Omega}, 
\label{eq: scaleHeight}
\end{equation}

\noindent where $\Omega$ is the bulk gas orbital frequency. For gas viscosity $\nu$ we employ the \citet{Shakura1973} prescription

\begin{equation}
\nu = \alpha c_{\rm s} H,
\label{eq: viscosity}
\end{equation}

\noindent where $\alpha$ is a dimensionless quantity smaller than unity. The appropriate value of $\alpha$ is unclear \citep{Kral2016BetaPic, Kral2016Magneto, Moor2019}, and so we leave it free for this analysis.

We assume the disc to be optically thin, such that its temperature is the blackbody temperature

 \begin{equation}
T_{\rm gas} = 278 \: {\rm K} \left( \frac{L_*}{L_\odot}\right)^{1/4} \left( \frac{r}{{\rm au}}\right)^{-1/2},
\label{eq: gasTemperature}
\end{equation}

\noindent where $L_*$ and $L_\odot$ are the star and solar luminosities respectively. The gas being optically thin means that relations between ${T_{\rm d}}$, $\beta$, $L_*$ and $r$ need not be modified for shadowing by the gas.

We can now asses the gas distribution. The surface density evolution equation for a thin, viscous gas disc \citep{LyndenBell1974, Pringle1981} with an additional term for mass input (e.g. \citealt{Kral2016BetaPic}) is

\begin{equation}
\frac{\partial \Sigma}{\partial t} = \frac{3}{r} \frac{\partial}{\partial r} \left[ r^{1/2} \frac{\partial}{\partial r} \left(\nu \Sigma r^{1/2} \right)\right] + \dot{\Sigma}_{\rm in},
\label{eq: surfaceDensityEvolutionEquationForThinDisc}
\end{equation}

\noindent where $t$ is time and $\dot{\Sigma}_{\rm in}$ is the surface density input rate. Gas is added to the disc by dust sublimation, which happens over a very narrow radius range; we therefore write

\begin{equation}
\dot{\Sigma}_{\rm in} = \frac{\dot{M}_{\rm in}}{2 \pi r_{\rm s}} \delta (r - r_{\rm s}),
\label{eq: surfaceDensityInputRate}
\end{equation}

\noindent where $\dot{M}_{\rm in}$ is the gas mass input rate at the sublimation radius $r_{\rm s}$, and $\delta (x)$ is the Dirac Delta Function.
 
Combining Equations \ref{eq: viscosity} and \ref{eq: gasTemperature} shows viscosity to be proportional to distance; to explicitly show this dependence, we define ${\nu \equiv V r}$ where $V$ is a parameter independent of $r$. We assume the disc to be in steady state, such that ${\partial \Sigma / \partial t = 0}$. Making these substitutions and expanding Equation \ref{eq: surfaceDensityEvolutionEquationForThinDisc} yields

\begin{equation}
2 r^2 \frac{\partial^2 \Sigma}{\partial r^2} + 7 r \frac{\partial \Sigma}{\partial r} + 3 \Sigma + \frac{r}{r_{\rm s}}\frac{\dot{M}_{\rm in}}{3 \pi V} \delta (r - r_{\rm s}) = 0.
\label{eq: surfaceDensityEvolutionSteadyState}
\end{equation}

\noindent This is a second order inhomogeneous linear differential equation, with the solution

\begin{equation}
\Sigma = \frac{C_1}{r} + \frac{C_2}{r^{3/2}} - \frac{\dot{M}_{\rm in}}{3 \pi V r} \left( 1 - \sqrt{\frac{r_{\rm s}}{r}} \right) \theta (r - r_{\rm s}).
\label{eq: secondOrderInhomogeneousSolution}
\end{equation}

\noindent Here $C_1$ and $C_2$ are constants to be determined, and $\theta(x)$ is the Heaviside step function. 

In general, an equation of the form of Equation \ref{eq: secondOrderInhomogeneousSolution} has its constants $C_1$ and $C_2$ determined by considering the homogeneous equivalent of Equation \ref{eq: surfaceDensityEvolutionSteadyState}, which is the case if ${\dot{M}_{\rm in} = 0}$. \citet{Armitage2010} states that for a thin disc extending to the surface of a slowly-rotating star, a simple solution to the homogeneous equation is 

\begin{equation}
\Sigma_{\rm h} = \frac{\dot{M}_{\rm acc}}{3 \pi V r} \left(1 - \sqrt{\frac{R_*}{r}}\right),
\label{eq: secondOrderHomogeneousSolution}
\end{equation}

\noindent where $\dot{M}_{\rm acc}$ is the accretion rate of gas onto the star, and $R_*$ is the star radius \citep{Armitage2010}. We define $C_1$ and $C_2$ from this, noting that Equation \ref{eq: secondOrderHomogeneousSolution} is a simplification and the actual gas properties close to the star would be complicated by various factors. Such a simplification is justified here because, as we later show, dust does not drift much farther inward than $r_{\rm s}$ and so does not encounter the inner region of the disc.

Substituting Equation \ref{eq: secondOrderHomogeneousSolution} into Equation \ref{eq: secondOrderInhomogeneousSolution} and rewriting the surface density in terms of that at $r_{\rm s}$ yields

\begin{multline}
\Sigma(r) = \Sigma(r_{\rm s}) \left( 1 - \sqrt{\frac{R_*}{r_{\rm s}}} \right)^{-1} \frac{r_{\rm s}}{r} 
\\
\times \left[ 1- \sqrt{\frac{R_*}{r}} - \theta (r - r_{\rm s})\frac{\dot{M}_{\rm in}}{\dot{M}_{\rm acc}} \left( 1 - \sqrt{\frac{r_{\rm s}}{r}} \right) \right],
\label{eq: gasSurfaceDensity}
\end{multline}

\noindent which can be compared to numerical results in \citet{Kral2016BetaPic} and \citet{Marino2020} at late times. By considering the scale height (Equation \ref{eq: scaleHeight}) we derive the midplane disc density as a function of radius as

\begin{multline}
\rho_{\rm gas}(r) = \rho_{\rm gas}(r_{\rm s}) \left( 1 - \sqrt{\frac{R_*}{r_{\rm s}}} \right)^{-1} \left(\frac{r_{\rm s}}{r}\right)^{9/4} 
\\
\times 
\left[ 1- \sqrt{\frac{R_*}{r}} - \theta (r - r_{\rm s}) \frac{\dot{M}_{\rm in}}{\dot{M}_{\rm acc}} \left( 1 - \sqrt{\frac{r_{\rm s}}{r}} \right) \right].
\label{eq: gasDensity}
\end{multline}

\begin{figure}
  \centering
   \includegraphics[width=7cm]{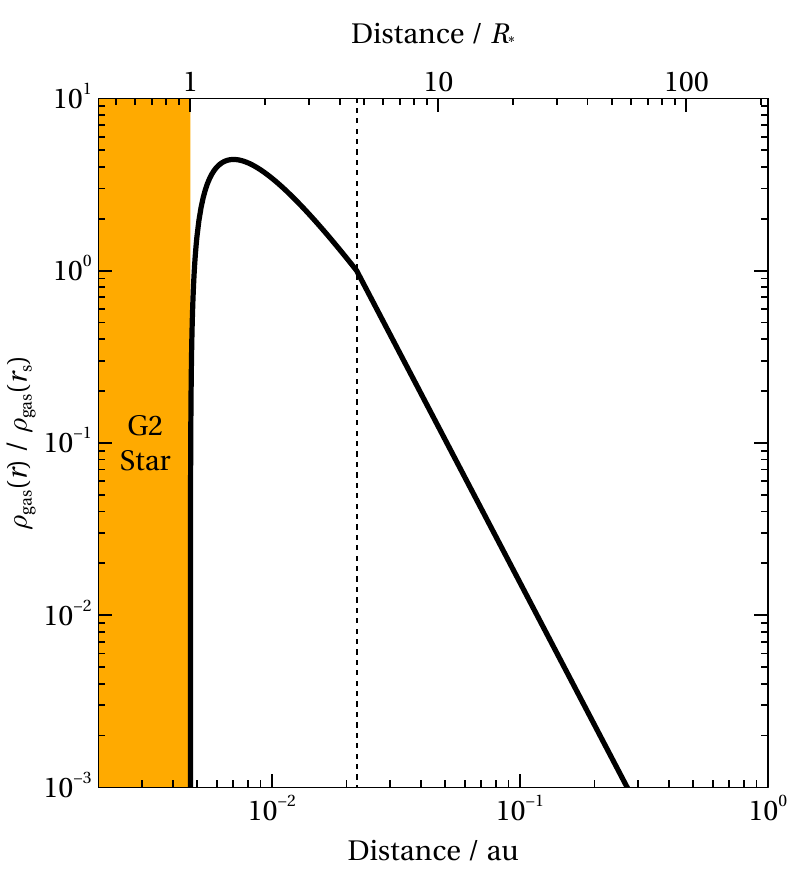}
   \caption{The midplane gas density relative to that at the dust sublimation radius, for steady-state gas around a G2 star. Here the gas mass input rate from sublimation is assumed to be similar to the gas accretion rate onto the star. The dashed line marks the sublimation radius $r_{\rm s}$ (taken to be ${0.022 \: {\rm au}}$), where gas mass is inputted.}
   \label{fig: gasDiscDensityProfile}
\end{figure}

\noindent The form of this density is shown on Figure \ref{fig: gasDiscDensityProfile}. Finally, integrating Equation \ref{eq: gasSurfaceDensity} yields the total gas mass within a radius $R$ as

\begin{multline}
M_{\rm gas}(r<R) = \sqrt{8 \pi^3} \rho_{\rm gas}(r_{\rm s}) H(r_{\rm s}) r_{\rm s} \left(1 - \sqrt{\frac{R_*}{r_{\rm s}}} \right)^{-1}
\\
 \times R \left[\left(1 - \sqrt{\frac{R_*}{R}}\right)^2 - \theta(R-r_{\rm s}) \frac{\dot{M}_{\rm in}}{\dot{M}_{\rm acc}} \left(1 - \sqrt{\frac{r_{\rm s}}{R}}\right)^2 \right].
\label{eq: gasMassWithinRadiusR}
\end{multline}

By examining gas density relative to that at a specific location in Equation \ref{eq: gasDensity}, we are delaying consideration of the viscosity parameter $\alpha$ and gas accretion rate $\dot{M}_{\rm acc}$ by absorbing them both into $\rho_{\rm gas}(r_{\rm s})$:

\begin{equation}
\rho_{\rm gas}(r_{\rm s}) = \frac{\dot{M}_{\rm acc}}{3 \sqrt{2 \pi^3} \alpha c_{\rm s} (r_{\rm s}) H^2 (r_{\rm s})} \left( 1 - \sqrt{\frac{R_*}{r_{\rm s}}} \right).
\label{eq: densityAtRsInTermsOfAccretionAndAlpha}
\end{equation}

\noindent Hence we may consider dust evolution for various values of $\rho_{\rm gas}(r_{\rm s})$ without picking values of $\alpha$ and $\dot{M}_{\rm acc}$; these values will be discussed in Section \ref{subsec: dustInflowRate}. However, Equation \ref{eq: gasDensity} does depend on the ratio of the gas mass input rate to the gas accretion rate onto the star, ${\dot{M}_{\rm in} / \dot{M}_{\rm acc}}$. A standard result for astrophysical discs is that almost all gas moves inwards, whilst a small fraction moves outwards in order to conserve angular momentum \citep{Pringle1981}. Hence we approximate ${\dot{M}_{\rm in} / \dot{M}_{\rm acc}} \approx 1$ for the remainder of this paper.

\subsubsection{Dust-gas interaction}
\label{subsec: dustGasInteractionImplementation}

\noindent We now consider the effect of the above gas disc on migrating dust. The grain equation of motion in the presence of gas is 

\begin{equation}
{\bf F} = {\bf F}_{\rm grav} + {\bf F}_{\rm rad} + {\bf F}_{\rm gas},
\label{eq: equationOfMotionGas}
\end{equation}

\noindent where ${\bf F}_{\rm gas}$ is the gas drag force. This force is related to ${\Delta \mathbf{v}}$, the grain velocity relative to the gas:

\begin{equation}
\Delta \mathbf{v} \equiv \mathbf{v}_{\rm d} - \mathbf{v}_{\rm gas}.
\label{eq: deltaV}
\end{equation}

\noindent We do not consider turbulent motion in the gas, hence the gas velocity ${\mathbf{v}_{\rm gas}}$ is the bulk velocity from Equation \ref{eq: gasVelocity}. Combining with Equation \ref{eq: dustVelocity}, for a grain on a circular orbit

\begin{equation}
\Delta \mathbf{v} = - {\bf v}_{\rm Kep} \left(\sqrt{1 - \eta} - \sqrt{1 - \beta} \right),
\label{eq: deltaVCircularOrbit}
\end{equation}

\noindent so dust on a circular orbit is slower than gas if ${\beta > \eta}$ (e.g. \citealt{Takeuchi2001}). In our parameter space ${\eta \sim 10^{-5}}$ to $10^{-4}$, with the smaller values closer to the star. A $\beta$ of $10^{-4}$ corresponds to a grain radius of several millimetres around a G2 star or several centimetres for an A0 star; for gas quantities considered in this paper, such large grains are not coupled to gas anyway. Hence in the parameter space where gas drag may be important ${\beta > \eta}$, and so dust moves more slowly than gas across the region of interest.

The drag formalism depends on grain size and ${\Delta \mathbf{v}}$. If the grain has radius ${s \ll \lambda}$, where $\lambda$ is the gas mean-free path, then drag is in the Epstein regime. The mean free path is

\begin{equation}
\lambda = \frac{\mu m_{\rm u}}{\sqrt{2} \sigma \rho_{\rm gas}},
\label{eq: meanFreePath}
\end{equation}

\noindent where ${\sigma}$ is the collisional cross-section of the gas. Even for ${s = 1 \: {\rm cm}}$, substituting ${\mu = 12.01}$ and ${\sigma \sim 10^{-16} \: {\rm cm}^2}$ for atomic carbon yields drag to be in the Epstein regime if ${\rho_{\rm gas} \ll 10^{-7} \: {\rm g \: cm^{-3}}}$. This is many orders of magnitude higher than gas densities used in this paper, so we assume that gas drag is always in the Epstein regime.

Epstein drag then depends on whether the grain is subsonic (${\Delta v \ll v_{\rm th}}$) or supersonic (${\Delta v \gg v_{\rm th}}$), where $v_{\rm th}$ is the mean thermal speed of gas molecules: 

 \begin{equation}
v_{\rm th} = \sqrt{\frac{8 k_{\rm B} T_{\rm gas}}{\pi \mu m_{\rm u}}}.
\label{eq: thermalSpeed}
\end{equation}

\noindent For subsonic grains ${F_{\rm gas} \propto v_{\rm th} \Delta v}$, whilst for supersonic grains ${F_{\rm gas} \propto \Delta v^2}$. We find very small grains to be supersonic, as are those on eccentric orbits. Larger grains on circular orbits are subsonic. To model both regimes and allow a smooth transition between them, we employ the formalism of \citet{Kwok1975} and define the gas drag force as

\begin{equation}
{\bf F}_{\rm gas} = -\frac{4\pi}{3} \rho_{\rm gas} s^2 \left(v_{\rm th}^2 + \Delta v^2 \right)^{1/2} \Delta \mathbf{v}.
\label{eq: gasDragForce}
\end{equation}

\subsubsection{Growth}
\label{subsec: growth}

\noindent As well as sublimating, grains embedded in gas can grow as material accretes onto them. Continuing with the prescription of \citet{Lebreton2013} and \citet{Lamy1974}, a grain embedded in gas of the same molecular weight will accrete particles at a rate

\begin{equation}
f_{\rm acc} = \frac{P_{\rm gas}}{\sqrt{2 \pi \mu m_{\rm u} k_{\rm B} T_{\rm gas}}},
\label{eq: accretionFlux}
\end{equation}

\noindent where $f_{\rm acc}$ is the flux of gas particles accreting onto the grain and $P_{\rm gas}$ is the gas pressure. Combining this with Equation \ref{eq: grainMassEvaporation} for sublimation, the grain mass changes at a rate

\begin{equation}
\frac{{\rm d}m}{{\rm d}t} = \gamma 4 \pi s^2 \left(f_{\rm acc} -  f_{\rm evap}\right) \mu m_{\rm u},
\label{eq: grainMassChange}
\end{equation}

\noindent and grain radius in a gas disc therefore evolves as

\begin{equation}
\frac{{\rm d}s}{{\rm d}t} = - \gamma \sqrt{\frac{k_{\rm B} T_{\rm d}}{2 \pi \mu m_{\rm u}}} \frac{\rho_{\rm eq} - \rho_{\rm gas} \sqrt{T_{\rm gas} / T_{\rm d}}}{\rho_{\rm d}}.
\label{eq: grainSizeEvolutionInGas}
\end{equation}

\noindent This is equivalent to Equation 17 of \citet{Lebreton2013}, noting that we do not require gas and dust to have the same temperature.


\subsection{Results in the presence of gas}
\label{subsec: evolutionWithGas}

\noindent Having implemented the gas model, we now describe the evolution of dust grains in the presence of gas.


\subsubsection{General grain evolution in the presence of gas}
\label{subsec: generalEvolutionWithGas}

\begin{figure*}
  \centering
   \includegraphics[width=16cm]{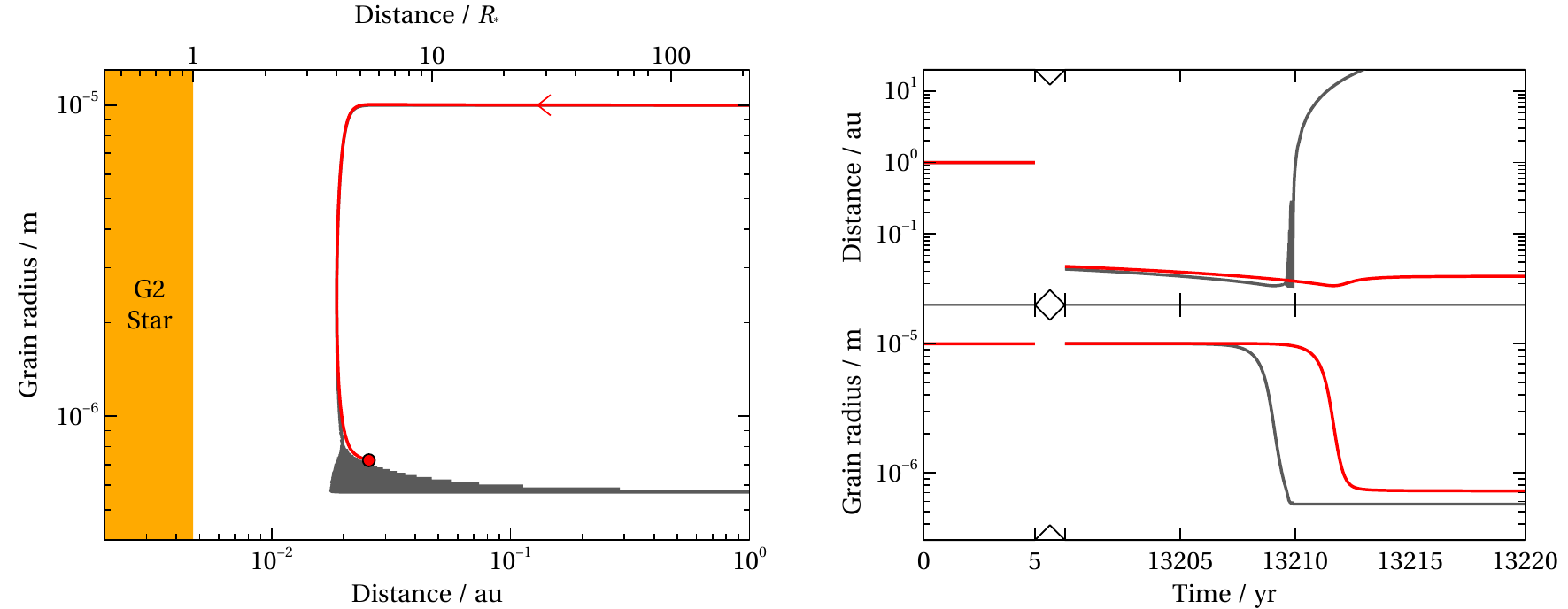}
   \caption{Evolution of an initially ${10 \: \mu {\rm m}}$ radius solid carbon grain orbiting a G2 star. Red and grey lines show evolution in the presence and absence of a gas disc, respectively. The gas density at ${r_{\rm s} = 0.022 \: {\rm au}}$ is ${\rho_{\rm gas}(r_{\rm s}) = 10^{-18} \: {\rm g \: cm}^{-3}}$, and the disc follows the density profile on Figure \ref{fig: gasDiscDensityProfile} (Equation \ref{eq: gasDensity}, with ${\dot{M}_{\rm in} = \dot{M}_{\rm acc}}$). The grain is initially too large to couple to gas, so its early evolution is identical to the gas-free case; the grain migrates inwards under PR-drag and begins to sublimate near the star. However, once the grain becomes sufficiently small it starts to couple to the gas, which draws the grain away from the star and protects it from further sublimation and eccentricity excitation. The grain is eventually trapped on a circular orbit just exterior to the sublimation region, where gas drag balances PR-drag. Note the break in the time axes on the right-hand plots, emphasising the speed of sublimation in comparison to drift.}
   \label{fig: evolutionWithContinuousGas}
\end{figure*}

\noindent Gas can trap grains larger than the blowout size exterior to the sublimation region, protecting them from further sublimation and so preventing their ejection from the system. Figure \ref{fig: evolutionWithContinuousGas} shows a simulation of an initially ${10 \: \mu {\rm m}}$ radius carbon grain orbiting a G2 star, in the presence of a gas disc with midplane density ${\rho_{\rm gas}(r_{\rm s}) = 10^{-18} \: {\rm g \: cm}^{-3}}$ at the sublimation distance of ${0.022 \: {\rm au}}$. Initially the grain is too large to be affected by gas, and its evolution is indistinguishable from the gas-free case on Figure \ref{fig: grainRadiusVsDistanceNoGas}; it migrates inwards under PR-drag, and begins to sublimate as it approaches the star. Here, however, the gas and gas-free behaviours diverge. As the grain sublimates it begins couple to the gas, and is drawn away from the star without its eccentricity being excited. This causes sublimation to slow. The grain continues to migrate outwards and eventually stalls, becoming indefinitely trapped on a circular orbit just exterior to the sublimation region. In the example on Figure \ref{fig: evolutionWithContinuousGas}, the grain is eventually trapped at ${0.026 \: {\rm au}}$ with radius ${0.72 \: \mu {\rm m}}$. As discussed below, this behaviour is qualitatively similar across a broad parameter space. Since dust orbits are rapidly circularised by gas, trapped grains can have $\beta$ values all the way up to ${\beta \approx 1}$ and still be on circular orbits.

\subsubsection{Criteria for dust trapping and growth}
\label{subsec: criteriaForGasTrapping}

\noindent Gas has two main effects on dust: drag and growth. Growth is simple; if gas is dense enough, sufficient material accretes onto the grain to offset that lost through sublimation. Equation \ref{eq: grainSizeEvolutionInGas} shows that a grain will grow if

\begin{equation}
\rho_{\rm gas} \sqrt{T_{\rm gas}} > \rho_{\rm eq} (T_{\rm d})\sqrt{T_{\rm d}}.
\label{eq: criterionForGrowth}
\end{equation}

Regarding gas drag, Equation \ref{eq: gasDragForce} shows that the drag force direction depends on the grain velocity relative to gas, and in our parameter space gas moves faster than dust (Section \ref{subsec: dustGasInteractionImplementation}). Gas drag therefore draws the grain away from the star, whilst PR-drag draws it towards the star. Hence a grain becomes trapped when gas drag balances PR-drag. 

We now calculate the gas density required to trap grains. The osculating semimajor axis $a$ of a grain subject to two-body gravity plus an additional, much smaller perturbing force instantaneously evolves as
 
\begin{equation}
\frac{{\rm d} a}{{\rm d} t} = 2 \frac{a^{3/2}}{\sqrt{GM_*(1-\beta)(1-e^2)}} \left[\bar{R} e \sin f + \bar{T} \left(1 + e \cos f \right) \right],
\label{eq: generalSemiMajorAxisChange}
\end{equation}
 
\noindent where $e$ and $f$ are grain eccentricity and true anomaly respectively, and $\bar{R}$ and $\bar{T}$ are the respective radial and tangential components of the perturbing acceleration \citep{Murray1999}. The trapped grains in our simulations have small eccentricities, since they are damped by both PR and gas drag. Therefore ${e = \bar{R} = 0}$, and gas drag causes the grain semimajor axis to expand at a rate
 
\begin{equation}
\frac{{\rm d} a}{{\rm d} t} = 2 \frac{a^{3/2}}{\sqrt{GM_*(1-\beta)}} \frac{\left|{\bf F}_{\rm gas}\right|}{m}
\label{eq: semimajorAxisChangeFromGasDrag}
\end{equation}
 
\noindent (noting that taking the magnitude of ${\bf F}_{\rm gas}$ is only valid if ${\beta > \eta}$, as satisfied for our parameter space). Similarly, PR-drag causes the semimajor axis to shrink at a rate
 
\begin{equation}
\frac{{\rm d} a}{{\rm d} t} = -2 \frac{\beta G M_*}{a c}.
\label{eq: semimajorAxisChangeFromPRDrag}
\end{equation}

\noindent Equating Equations \ref{eq: semimajorAxisChangeFromGasDrag} and \ref{eq: semimajorAxisChangeFromPRDrag} and substituting ${\bf F}_{\rm gas}$ from Equation \ref{eq: gasDragForce}, a grain will be trapped if the local gas density satisfies

\begin{equation}
\rho_{\rm gas} = \rho_{\rm d} \frac{s\beta \sqrt{1-\beta}}{c} \frac{\left( G M_* \right)^{3/2}}{r^{5/2} \sqrt{v_{\rm th}^2 + \Delta v^2} \Delta v}.
\label{eq: gasDensityRequiredForTrapping}
\end{equation}

\begin{figure*}
  \centering
   \includegraphics[width=16cm]{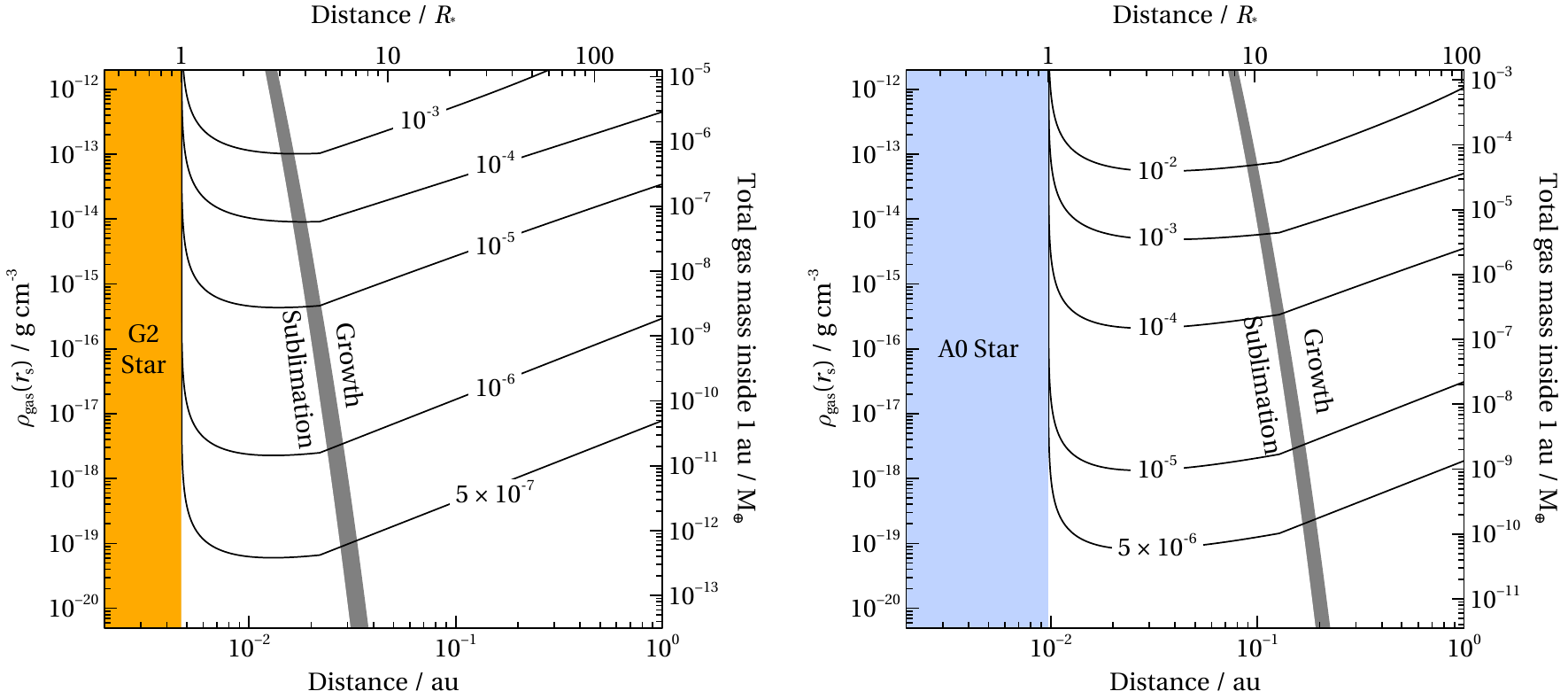}
   \caption{Gas quantities required to trap non-eccentric, solid carbon dust grains at different stellar distances. The left plot shows a G2 star for which the sublimation radius is ${r_{\rm s} = 0.022 \: {\rm au}}$, and the right plot shows an A0 star for which ${r_{\rm s} = 0.13 \: {\rm au}}$. Vertical axes show the gas density at $r_{\rm s}$ required to trap grains, and the corresponding total gas mass within ${1 \: {\rm au}}$ (calculated using Equation \ref{eq: gasMassWithinRadiusR}). Contours show different grain radii (in metres), and grey bands denote the boundary between grain sublimation and growth (shown as bands rather than lines because values differ slightly for different grain sizes). For example, a ${10^{-6} \: {\rm m}}$ radius grain can be trapped at ${0.1 \:{\rm au}}$ around a G2 star if the gas density at the sublimation radius is ${10^{-17} \: {\rm g \: cm}^{-3}}$, corresponding to a total gas mass inside ${1 \: {\rm au}}$ of ${10^{-10} \: {\rm M}_\oplus}$. Note that the left-hand vertical axes show gas density at $r_{\rm s}$, rather than local density; the local density at each radius can be found from Figure \ref{fig: gasDiscDensityProfile}. The plots show the analytic arguments in Section \ref{subsec: criteriaForGasTrapping} and agree with simulation results; provided ${\rho_{\rm gas}(r_{\rm s}) \gtrsim 10^{-19} \: {\rm g \: cm}^{-3}}$, grains around the blowout size can be trapped for very long periods just exterior to the sublimation radius of G2 and A0 stars (the blowout radius for solid carbon is $0.5$ and ${5 \: \mu {\rm m}}$ for G2 and A0 stars, respectively). Note that the total gas mass for a given ${\rho_{\rm gas}(r_{\rm s})}$ is larger for A0 stars than G2 stars, because the sublimation radius is farther away from the former type. Grains larger than those shown will not be trapped, because they experience an inwards drag force from the gas. For a given gas density, all grains eventually lie at the corresponding point in the grey region regardless of their initial size, due to the effect described in Section \ref{subsec: criteriaForGasTrapping}.}
   \label{fig: trappingAndGrowthInContinuousGasDensity}
\end{figure*}

\noindent Finally, Equation \ref{eq: gasDensity} relates the local gas density at radius $r$ to that at the sublimation radius, ${\rho_{\rm gas} (r_{\rm s})}$. Figure \ref{fig: trappingAndGrowthInContinuousGasDensity} shows the gas density at the sublimation radius required to trap grains of different sizes at different distances, and the distances at which grains will sublimate or grow. Gas can trap grains just exterior to the sublimation region if the gas density at the sublimation radius is greater than ${10^{-19} \: {\rm g \: cm^{-3}}}$ for both G2 and A0 stars; these densities correspond to total gas masses inside ${1 \: {\rm au}}$ of ${10^{-12}}$ and ${10^{-10} \: {\rm M_\oplus}}$, respectively.

Both PR and gas drag are more effective for smaller grains. Insight into their relative strengths is found by considering a grain on a circular orbit, and approximating ${\beta \propto 1/s}$. Taking the gas density required for trapping from Equation \ref{eq: gasDensityRequiredForTrapping}, and substituting ${\Delta v}$ from Equation \ref{eq: deltaVCircularOrbit} (with ${\eta \ll 1}$), shows that $\rho_{\rm gas}$ scales with either ${\sqrt{1-\beta} / (1 - \sqrt{1-\beta})}$ or ${\sqrt{1-\beta} / (1 - \sqrt{1-\beta})^2}$. Hence gas trapping becomes more effective than PR-drag at smaller grain radii (higher $\beta$), and so less gas is required to trap smaller grains than larger ones (Figure \ref{fig: trappingAndGrowthInContinuousGasDensity}).

Since gas density falls rapidly with distance (Figure \ref{fig: gasDiscDensityProfile}), the total gas mass required to instantaneously trap grains significantly increases farther from the star. However, grains cannot be trapped interior to the sublimation region (left of the grey bands on Figure \ref{fig: trappingAndGrowthInContinuousGasDensity}), because they would continue to sublimate; as they became smaller, gas drag would become more effective and eventually pull the grains outwards. Likewise, grains instantaneously trapped exterior to the sublimation region (right of the grey bands on Figure \ref{fig: trappingAndGrowthInContinuousGasDensity}) would grow and eventually decouple from the gas, then migrate inwards under PR-drag. Migrating grains would then encounter a higher gas density closer to the star, and become trapped again. Hence these processes cause grains to migrate towards the grey shaded regions of Figure \ref{fig: trappingAndGrowthInContinuousGasDensity}, where they neither grow nor sublimate, and become trapped indefinitely. Of course, the actual time they remain trapped for depends on unmodelled physical processes, as discussed in Section \ref{subsec: lifetimeOfTrappedGrains}. 

\begin{figure}
  \centering
   \includegraphics[width=7cm]{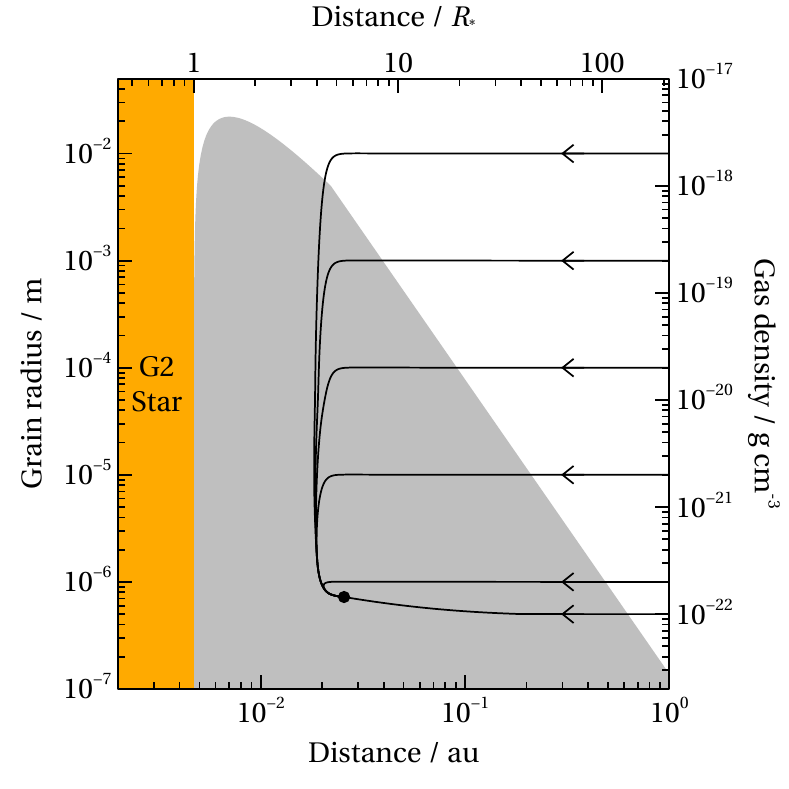}
   \caption{For a given gas density, the final size and distance of a trapped grain is independent of its initial parameters. Shown is the evolution of solid carbon grains of various initial sizes around a G2 star, in the presence of a gas disc with density ${10^{-18} \: {\rm g \: cm}^{-3}}$ at ${0.022 \: {\rm au}}$. Black lines show the evolving grain radii (left-hand vertical axis) versus distance. The shaded grey area shows the local gas density (right-hand vertical axis). The black circle shows the final grain parameters at the end of the simulations, which are independent of initial grain size.}
   \label{fig: convergingGrainParametersForSingleGasDensity}
\end{figure}

\begin{figure}
  \centering
   \includegraphics[width=7cm]{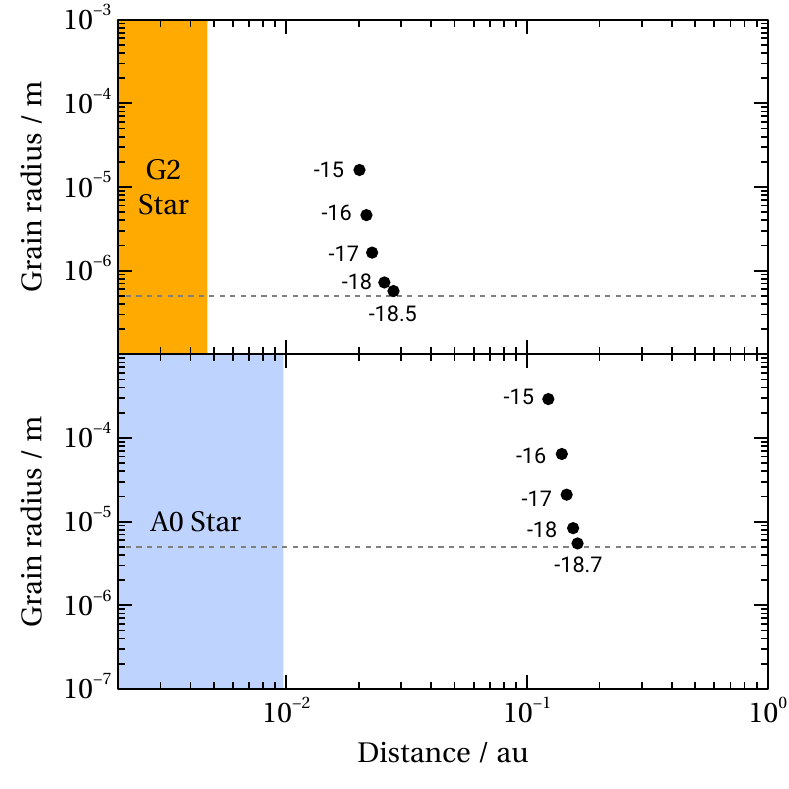}
   \caption{The final radii and distances of trapped grain populations from simulations with different gas densities. Labels show ${\log_{10}[\rho_{\rm gas}(r_{\rm s}) / ({\rm g \; cm^{-3}})]}$, where ${\rho_{\rm gas}(r_{\rm s})}$ is the gas density at the sublimation radius. Dashed lines show grain blowout radii; grains smaller than this cannot be trapped by gas. Note that the trapped grain parameters are independent of initial grain sizes and depend only on gas quantity, as shown on Figure \ref{fig: convergingGrainParametersForSingleGasDensity}.}
   \label{fig: finalGrainParametersForDifferentGasDensities}
\end{figure}

For a given gas density the sizes and distances of trapped grains converge, such that the eventual trapping location and grain size is independent of initial grain size. This is shown on Figure \ref{fig: convergingGrainParametersForSingleGasDensity}. This means that, for a given star type, the eventual size and location of dust depends only on the amount of gas, and not on the initial grain parameters. Figure \ref{fig: finalGrainParametersForDifferentGasDensities} shows the final properties of trapped grains for different gas densities from simulations; lower gas quantities result in smaller grains being trapped farther from the star, whilst higher gas quantities result in larger grains being trapped closer to the star.

\subsection{Behaviour of sub-blowout grains}
\label{subsec: behaviourOfSubblowoutGrains}

\noindent Grains smaller than the blowout size could potentially be created in collisions between trapped or migrating dust (see Section \ref{sec: dustLifetimeAndObservationalComparison}). For solid carbon grains the blowout radii are just below ${0.5}$ and ${5 \: \mu {\rm m}}$ for G2 and A0 stars, respectively. However, such unbound grains cannot be trapped by this mechanism, regardless of gas quantity. This is because gas drag is much weaker than radiation pressure in our parameter space, so cannot prevent high $\beta$ grains from being ejected. Dramatically increasing gas densities does not help; in this regime grains rapidly grow, so even if sub-blowout grains were trapped, they would quickly grow larger than the blowout size.

However, sub-blowout grains can be slowed by gas and take longer to escape the system. In the absence of gas, a small grain with ${\beta > 1}$ will follow an anomalous hyperbolic trajectory, where it is accelerated out of the system by radiation pressure. Its speed continuously increases and tends towards a value at infinity, $v_\infty$. This speed can be derived from energy conservation; since ${v^2 - 2 G M_* (1-\beta) / r}$ is constant, if a trapped grain on a circular orbit at distance $r_0$ with ${\beta_0 < 1}$ releases an unbound grain with ${\beta > 1}$, the unbound grain will accelerate away from the star with a speed tending towards

\begin{equation}
v_\infty = \sqrt{\frac{G M_*}{r_0}} \sqrt{2 \beta - \beta_0 - 1}.
\label{eq: velocityAtInfinity}
\end{equation}

\begin{figure}
  \centering
   \includegraphics[width=7cm]{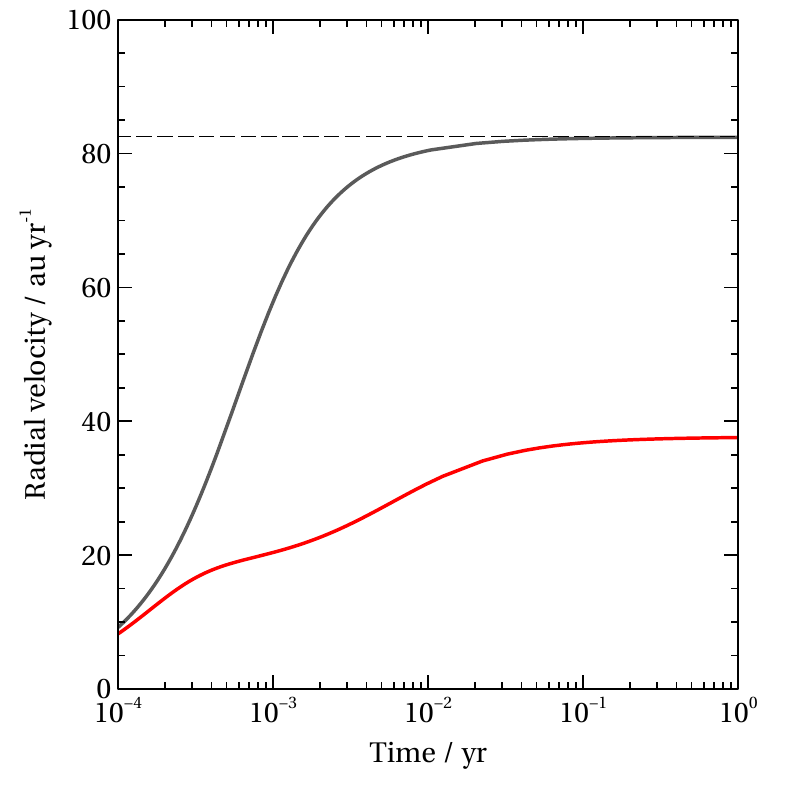}
   \caption{The gas cannot trap unbound grains, but it can substantially slow their escape. Such grains could potentially be created in collisions between trapped or migrating dust. This plot shows the radial velocity over time of an unbound grain from simulations with (red) and without (grey) a gas disc. The dashed line shows the theoretical velocity at infinity in the absence of gas (Equation \ref{eq: velocityAtInfinity}); the final velocity of the grain that interacted with gas is significantly lower than this, even once it is well-clear of the gas. This particular grain is ${0.05 \; \mu {\rm m}}$ in radius, and was released at ${0.04 \; {\rm au}}$ from a G2 star (just exterior to the sublimation radius). Its initial velocity was equal to that of a ${1 \; \mu {\rm m}}$ grain on a circular orbit at ${0.04 \; {\rm au}}$, and the gas disc has a density at the sublimation radius of ${\rho_{\rm gas}(r_{\rm s}) = 10^{-15} \; {\rm g \; cm^{-3}}}$.}
   \label{fig: unboundGrainTrajectories}
\end{figure}

\noindent This behaviour is shown by the solid grey line on Figure \ref{fig: unboundGrainTrajectories}. If gas is present, however, the above argument no longer holds; since gas drag is a non-conservative force, the gas removes kinetic energy from the escaping grain. The grain is therefore not only slower as it passes through the gas, but also its speed is reduced compared to the gas-free case even once it is well-clear of the gas. So an unbound grain that has passed through a gas disc will never reach the speed at infinity from Equation \ref{eq: velocityAtInfinity}, even though it will quickly leave the gas region. This is shown by the red line on Figure \ref{fig: unboundGrainTrajectories}. The presence of gas can therefore significantly slow the exit of unbound grains from the system, even if it cannot trap them indefinitely.

\vspace{1cm}


\section{Dust lifetime and comparison to observations}
\label{sec: dustLifetimeAndObservationalComparison}

\noindent We have shown that modest quantities of gas can trap dust close to stars. Here we discuss the lifetime of trapped grains, the required dust inflow rate, and whether our model can explain NIR excesses.

\subsection{Lifetime of trapped grains}
\label{subsec: lifetimeOfTrappedGrains}

\noindent In our simple model, dust trapped in gas remains so indefinitely. In reality additional processing is expected to occur which limits the lifetime of trapped dust. Two potential mechanisms are collisions between trapped grains, and transportation by viscously spreading gas.

\subsubsection{Collisions between trapped grains}
\label{subsec: lifetimeFromCollisions}

Destructive collisions may limit dust lifetime. However, the efficiency of this process is unclear; whilst collisions may be frequent close to the star, eccentricities could be so damped that collisions are not destructive. Here we consider collisions between grains of a single size confined to a narrow ring, which is the eventual state of trapped grains (Figure \ref{fig: convergingGrainParametersForSingleGasDensity}). For small grains the average collision speed $v_{\rm col}$ is

\begin{equation}
v_{\rm col} = f(e, I) v_{\rm Kep} \sqrt{1-\beta},
\label{eq: collisionSpeed}
\end{equation}

\noindent where ${f(e, I) = \sqrt{1.25e^2 + I^2}}$ and $e$ and $I$ are the average grain eccentricity and inclination, respectively \citep{Wyatt2002}. The collision rate is ${n \sigma_{\rm col} v_{\rm col}}$, where $n$ is the grain number density and ${\sigma_{\rm col} = 4 \pi s^2}$ is the interaction cross-section. Hence for a ring of radius $r$, width $\Delta r$ and height ${2 r I}$, the time between collisions is

\begin{equation}
t_{\rm col} = \frac{4 \pi \rho_{\rm d} s r^{2.5} \Delta r}{3 M_{\rm d} \sqrt{G M_* (1-\beta)}}\frac{I}{f(e,I)}.
\label{eq: collisionTime}
\end{equation}

\noindent Taking ${I \approx e}$, ${r = r_{\rm s}}$ and ${\Delta r / r = 0.1}$, the collision time is ${0.07 \: {\rm yr}}$ for ${10^{-11} \; {\rm M}_\oplus}$ of ${1 \: {\rm \mu m}}$ dust trapped around a G2 star, and ${2 \: {\rm yr}}$ for ${10^{-9} \; {\rm M}_\oplus}$ of ${10 \: {\rm \mu m}}$ dust trapped around an A0 star (these dust masses are required to explain NIR observations in \citealt{Kirchschlager2017}). Hence the collision rate of trapped material is very high.

However, these frequent collisions do not necessarily preclude long dust lifetimes, since gas drag can substantially damp dust eccentricities (and hence collision speeds). From \citet{Wyatt2002}, a collision between equal-size grains will be catastrophic (i.e. the largest surviving grain has less than half the mass of each original grain) if ${v_{\rm col}^2 > 2Q_{\rm D}^*}$, where $Q_{\rm D}^*$ quantifies the size-dependent material strength. We could not locate literature $Q_{\rm D}^*$ values for solid graphite, so approximate them as those for basalt from Figure 6 in \citet{Wyatt2002}. Collisions are therefore catastrophic if $v_{\rm col}$ exceeds ${100 \; {\rm ms^{-1}}}$ for ${1 \: {\rm \mu m}}$ grains, and ${60 \; {\rm ms^{-1}}}$ for ${10 \: {\rm \mu m}}$ grains. From Equation \ref{eq: collisionSpeed}, these speeds correspond to eccentricities of order ${10^{-4} - 10^{-3}}$ for trapped material around G2 and A0 stars. Hence the eccentricities of trapped grains must be very small in order to survive frequent collisions.

Whether such low eccentricities are possible depends on the gas disc. An axisymmetric, non-turbulent disc very effectively damps eccentricities;  in our simulations of discs with ${\rho_{\rm gas}(r_{\rm s}) = 10^{-17} \; {\rm g \; cm^{-3}}}$, even grains with initial eccentricities of $0.9$ and pericentres at ${1 \; {\rm au}}$ have eccentricities damped to ${\lesssim 10^{-3}}$ by the time they migrate to the sublimation region. However, realistic discs also have turbulence parameterised by $\alpha$, which would drive up dust eccentricities \citep{Ida2008}. The magnitude of this effect also depends on the Stokes number ${\rm St}$, which is roughly a measure of how many orbital periods it takes for gas to significantly affect the grain orbit; ${{\rm St} = \Omega m \Delta v / |{\bf F}_{\rm gas}|}$ \citep{Armitage2010}. For trapped grains with parameters from Figure \ref{fig: finalGrainParametersForDifferentGasDensities}, Stokes numbers are ${> 800}$ for dust trapped around G2 stars and ${> 1700}$ for that around A0 stars (despite large Stokes numbers, gas affects dust over small timescales because the grain orbital periods are so short). In this regime, gas turbulence imparts collision speeds between equal-size grains of

\begin{equation}
v_{\rm col} = c_{\rm s} \sqrt{\frac{3 \alpha}{1 + {\rm St}}}
\label{eq: collisionSpeedFromTurbulence}
\end{equation}

\noindent (\citealt{Ormel2008}, their Equation 10). This is a measure of the minimum possible average collision velocity, because gas drag cannot reduce dust eccentricity below that imparted by turbulence. The sound speed at the trapping distance is ${1100 \; {\rm ms}^{-1}}$, so turbulence in high viscosity discs ($\alpha \sim 1$) could bring trapped grains to collision velocities of ${70 \; {\rm ms}^{-1}}$ around G2 stars and ${50 \; {\rm ms}^{-1}}$ around A0 stars, which are at the lower end of the estimated catastrophic collision speeds. So even for very high disc viscosities, eccentricity damping by gas may prevent collisions between trapped grains from being catastrophic.

Even if collisions are not catastrophic then trapped grains do not necessarily survive collisions indefinitely, since mass may still be lost through lower-speed cratering collisions. Grains undergoing such collisions would erode until they reached the blowout size, at which point they would leave the system. We now estimate the collisional lifetime of such an eroding grain. Following a collision between two grains, each of mass $m_0$, the largest remnant of each would have mass $f m_0$. Here 

\begin{equation}
f \approx 1 - \frac{v_{\rm col}^2}{4 Q_{\rm D}^*}
\label{eq: f}
\end{equation}

\noindent for cratering collisions in the strength regime \citep{Wyatt2002}, where $v_{\rm col}$ is set by disc turbulence (Equation \ref{eq: collisionSpeedFromTurbulence}). If trapped grains are only just larger than the blowout size then $f$ will not change significantly as the grain becomes smaller, so after $N$ collisions the grain radius is approximately ${s_0 f^{N/3}}$, where $s_0$ is the radius of the original trapped grain. The number of collisions that a trapped grain can survive is therefore

\begin{equation}
N \approx \frac{3 \log(s_{\rm bl} / s_0) }{ \log(f)},
\label{eq: numberOfCollisionsGrainCanSurvive}
\end{equation}

\noindent where $s_{\rm bl}$ is the blowout radius. Note that $f$ is close to unity for our parameter space, and so grains can survive many collisions; for example, for ${10 \; \mu {\rm m}}$ dust trapped around an A0 star with ${\alpha = 0.1}$, ${f > 0.97}$ and ${N > 70}$. The collisional lifetime is then ${\tau_{\rm col} \approx N t_{\rm col}}$, and roughly scales as $\alpha^{-1}$; a ${1 \; \mu {\rm m}}$ grain trapped around a G2 star has a collisional lifetime ${\tau_{\rm col} \approx 0.7 \; {\rm yr} / \alpha}$, and a ${10 \; \mu {\rm m}}$ grain trapped around an A0 star has a collisional lifetime ${\tau_{\rm col} \approx 20 \; {\rm yr} / \alpha}$.

A further effect is that collisional erosion could be offset by grain growth via gas accretion. As the grain shrinks it is pulled away from the star by gas, and therefore cools. Since sublimation depends much more strongly on temperature than accretion does, the grain grows. The radius of a trapped grain initially erodes at a rate ${|\dot{s}| = s_0 (1-f^{1/3}) / t_{\rm col}}$, and equating this to the rate of growth from gas accretion near the sublimation radius (Equation \ref{eq: grainSizeEvolutionInGas}) yields that for both star types collisional erosion is offset by gas accretion for all $\alpha$  provided ${\rho_{\rm gas}(r_{\rm s}) \gtrsim 10^{-16} \; {\rm g \; cm^{-3}}}$. In this case trapped grain lifetimes are not limited by collisions.

The conclusion is that, whilst collisions between trapped grains are frequent, efficient eccentricity damping by gas means that these grains are likely to survive multiple collisions. In this case the collisional lifetime is roughly proportional to $\alpha^{-1}$, and can be hundreds of years if ${\alpha < 10^{-2}}$. Alternatively if the gas density is greater than ${10^{-16} \; {\rm g \; cm^{-3}}}$ then grains can survive collisions indefinitely for any $\alpha$, as accretion can offset collisional erosion. Note that different assumptions about dust composition would change $Q_{\rm D}^*$ and potentially alter these specific conclusions, as the collisional lifetime roughly scales as ${Q_{\rm D}^* / \alpha}$.

\subsubsection{Viscous spreading}
\label{subsec: lifetimeFromViscousSpreading}

Trapped dust lifetimes could also be limited by the motion of gas. Assuming the gas to be atomic carbon, which feels little radiation pressure \citep{Fernandez2006}, radial gas motion is likely dominated by viscous spreading. Gas created by sublimating dust would spread away from the sublimation region, with the majority moving inwards and eventually accreting onto the star \citep{Pringle1981}. Gas is expected to drag trapped grains inwards with it, such that the dust would eventually sublimate and potentially escape. Hence another estimate of the trapped dust lifetime is the gas viscous timescale, which characterises how quickly gas spreads. For gas at distance $r$ the viscous timescale is ${\tau_{\nu} = r^2 / (3 \nu)}$ (e.g. \citealt{Papaloizou1999}), and so

\begin{equation}
\tau_{\nu} \sim \frac{\sqrt{G M_* r}\mu m_{\rm u}}{3 \alpha k_{\rm B} T_{\rm gas}}.
\label{eq: viscousTimescale}
\end{equation}

\noindent Substituting appropriate values, the viscous timescale at the sublimation radius is ${5 \: {\rm yr} / \alpha}$ for a G2 star and ${20 \: {\rm yr} / \alpha}$ for an A0 star. The viscous lifetime is therefore comparable to the collisional lifetime for trapped dust around A0 stars, and an order of magnitude longer than the collisional lifetime for grains around G2 stars. As discussed above, if the gas density at the sublimation radius is comparable to ${10^{-16} \; {\rm g \; cm^{-3}}}$ or greater, then accretion onto grains can offset collisional erosion; in this case the trapped dust lifetime would be set by viscous spreading of gas, rather than collisions.

\subsection{Required dust inflow rate}
\label{subsec: dustInflowRate}

\noindent We now consider the dust supply required to maintain a trapped population. We make no assumption about the source, which could be a collisional debris disc, comets, or something else, provided that dust originates exterior to the sublimation region and migrates inwards. The only constraint is on dust inflow rate, as set by two requirements: the need to replace grains as they are lost, and to replenish gas as it accretes onto the star.

As described in Section \ref{subsec: lifetimeFromCollisions}, trapped grains likely undergo cratering collisions. These collisions release much smaller grains, which must have radii smaller than ${s_0 (1-f)^{1/3}}$. These ejecta would be below the blowout size in our parameter space and therefore escape, resulting in mass loss from the system. Furthermore, over the collision lifetime ${\tau_{\rm col}}$ the grain erodes down to the blowout size, at which point it also escapes. The mass loss rate through collisions in the whole trapped population is therefore

\begin{equation}
\dot{M}_{\rm col} = \frac{M_d}{\tau_{\rm col}},
\label{eq: massInflowRateToOffsetCollisions}
\end{equation}

\noindent where $\tau_{\rm col}$ was derived in Section \ref{subsec: lifetimeFromCollisions}. The mass loss rate from collisions is therefore ${2 \times 10^{-11} \; {\rm M_\oplus} \; {\rm yr}^{-1} \alpha}$ for ${10^{-11} \: {\rm M_\oplus}}$ of ${1 \; \mu {\rm m}}$ grains trapped around a G2 star, and ${5 \times 10^{-11} \; {\rm M_\oplus} \; {\rm yr}^{-1} \alpha}$ for ${10^{-9} \: {\rm M_\oplus}}$ of ${10 \; \mu {\rm m}}$ grains trapped around an A0 star.

Dust can also be lost through viscous gas spreading, as described in Section \ref{subsec: lifetimeFromViscousSpreading}. If trapped grains survive for a viscous timescale $\tau_\nu$, then viscous spreading causes mass loss at a rate ${M_{\rm d} / \tau_\nu}$. Assuming typical parameters, this mass loss rate is ${2 \times 10^{-12} \: {\rm M_\oplus \: yr^{-1}}\alpha}$ for trapped grains around a G2 star and ${5 \times 10^{-11} \: {\rm M_\oplus \: yr^{-1}}\alpha}$ for those around an A0 star.

Finally, mass is also lost as gas itself accretes onto the star. This occurs at a rate ${\dot{M}_{\rm acc}}$, approximately equal to the gas input rate ${\dot{M}_{\rm in}}$. Provided that source grains are larger than trapped grains, most inflowing mass sublimates into gas (Figure \ref{fig: convergingGrainParametersForSingleGasDensity}). Rearranging Equation \ref{eq: densityAtRsInTermsOfAccretionAndAlpha}, gas accretes onto the star at a rate 

\begin{equation}
\dot{M}_{\rm acc} = 3\sqrt{2 \pi^3} \alpha \rho_{\rm gas}(r_{\rm s}) c_{\rm s}(r_{\rm s}) H^2 (r_{\rm s}) \left(1 - \sqrt{\frac{R_*}{r_{\rm s}}}\right)^{-1}.
\label{eq: massInflowRateToMaintainDisc}
\end{equation}

\noindent This equates to ${9 \times 10^{-12} \; {\rm M_\oplus} \; {\rm yr}^{-1} \alpha [\rho_{\rm gas}(r_{\rm s}) / 10^{-16} \; {\rm g \; cm^{-3}}]}$ for dust trapped around a G2 star, and ${6 \times 10^{-10} \; {\rm M_\oplus} \; {\rm yr}^{-1} \alpha [\rho_{\rm gas}(r_{\rm s}) / 10^{-16} \; {\rm g \; cm^{-3}}]}$ for that around an A0 star.

Combining all three mass loss processes, sustaining a gas trap requires a mass input rate of

\begin{equation}
\dot{M} = 2 \times 10^{-11} \; {\rm M_\oplus} \; {\rm yr}^{-1} \alpha \left\{ 1 + 0.5 \left[\frac{\rho_{\rm gas}(r_{\rm s})}{10^{-16} \; {\rm g \; cm^{-3}}} \right] \right\}
\label{eq: massLossRateG2Star}
\end{equation}

\noindent for ${10^{-11} \: {\rm M_\oplus}}$ of ${1 \; \mu {\rm m}}$ grains trapped around a G2 star, and 

\begin{equation}
\dot{M} = 1 \times 10^{-10} \; {\rm M_\oplus} \; {\rm yr}^{-1} \alpha \left\{ 1 + 6 \left[\frac{\rho_{\rm gas}(r_{\rm s})}{10^{-16} \; {\rm g \; cm^{-3}}} \right] \right\}
\label{eq: massLossRateA0Star}
\end{equation}

\noindent for ${10^{-9} \: {\rm M_\oplus}}$ of ${10 \; \mu {\rm m}}$ grains trapped around an A0 star. These mass inflows are small, equivalent to a ${\sim 1 \; {\rm km}}$ radius comet fully sublimating per year if ${\alpha \sim 1}$. However, unless $\alpha$ is small they still exceed the dust mass inflow into the inner Solar System, estimated as ${5 \times 10^{-14} \; {\rm M_\oplus \: yr^{-1}}}$ \citep{Grun1985}. These low mass inflows imply that the mechanism may be common across a diverse range of stellar systems, but is not necessarily ubiquitous unless $\alpha$ is small.

It should also be noted that gas trapping is self-limiting. A significant dust inflow increase, for example after a major collision in an external debris disc, would lead to increased sublimation and a greater gas density. As this denser gas spread it would trap incoming grains before they reached the sublimation radius, preventing further sublimation; gas density would therefore reduce again as it accreted onto the star. The reduced disc would be unable to trap grains outside the sublimation radius, so sublimation could resume and the disc grow again. This effect may limit the maximum possible gas mass.

\subsection{Dust mass enhancement factor}
\label{subsec: dustMassEnhancementFactor}

\noindent The presence of gas would enhance the dust quantity close to the star, compared to that expected from inward-migrating grains alone. Such gas would reduce the dust inflow rate required to maintain any hot dust population, regardless of the origins of the grains and gas. We now calculate the factor by which the dust quantity is increased by gas trapping, assuming that dust is created at an external source and that it migrates inwards under PR-drag.

If no gas trap operates, then grains of radius $s$ acting under PR-drag will migrate from distance $r_2$ to $r_1$ in a time

\begin{equation}
\Delta t(s) = \frac{c}{4 \beta (s) G M_*} \left(r_2^2 - r_1^2 \right).
\label{eq: prMigrationTime}
\end{equation}

\noindent If these grains are continuously created by an external source at a rate ${\dot{M}(s)}$, then the total mass in grains of this size between distances $r_1$ and $r_2$ is 

\begin{equation}
M(s) = \dot{M}(s) \Delta t (s).
\label{eq: inflowingMassInAnnulus}
\end{equation}

\noindent We assume that migrating dust follows a size distribution with number density ${\propto s^{-p}}$ where ${p < 4}$, such that most of the dust mass is contained in the largest grains. Larger grains also take longer to migrate, since in this regime ${\beta \propto s^{-1}}$. This means that the total migrating dust mass between $r_1$ and $r_2$ is approximately that in the largest grains with radii $s_{\rm max}$, i.e. ${M_{\rm pr} \approx \dot{M} \Delta t (s_{\rm max})}$, where $\dot{M}$ is the total mass inflow rate across all grain sizes.

Now Equations \ref{eq: massLossRateG2Star} and \ref{eq: massLossRateA0Star} give the mass input rates required to sustain gas traps containing ${10^{-11} \; {\rm M_\oplus}}$ of ${1 \; \mu {\rm m}}$ grains around a G2 star, and ${10^{-9} \; {\rm M_\oplus}}$ of ${10 \; \mu {\rm m}}$ grains around an A0 star, respectively. By comparing these trapped dust masses to those that would arise in the same region from the same mass inflow rates under PR-drag alone, we can calculate the dust mass enhancement factor ${X \equiv M_{\rm d} / M_{\rm pr}}$ caused by the gas trap. We assume the trapped dust occupies a narrow disc of central radius $r_{\rm s}$ and width ${\Delta r / r_{\rm s} = 0.1}$, and as an upper limit we assume that the size of the largest migrating grain equals that of the trapped grains. In this case the presence of gas enhances the dust mass close to the star by a factor

\begin{equation}
X = 5 / \alpha \left\{ 1 + 0.5 \left[\frac{\rho_{\rm gas}(r_{\rm s})}{10^{-16} \; {\rm g \; cm^{-3}}} \right] \right\}^{-1}
\label{eq: dustMassEnhancementFactorG2Star}
\end{equation}

\noindent for ${10^{-11} {\rm M_\oplus}}$ of ${1 \; \mu {\rm m}}$ grains trapped around a G2 star, and

\begin{equation}
X = 7 / \alpha \left\{ 1 + 6 \left[\frac{\rho_{\rm gas}(r_{\rm s})}{10^{-16} \; {\rm g \; cm^{-3}}} \right] \right\}^{-1}
\label{eq: dustMassEnhancementFactorA0Star}
\end{equation}

\noindent for ${10^{-9} \; {\rm M_\oplus}}$ of ${10 \; \mu {\rm m}}$ grains trapped around an A0 star. These are upper limits because we have assumed that the maximum migrating grain radii are $1$ and ${10 \; \mu {\rm m}}$ for G2 and A0 stars respectively; if the largest migrating grains are larger than this with radius $s_{\rm max}$, then the above enhancement factors will decrease as ${X \propto s_{\rm max}^{-1}}$. If ${p > 3}$ then the total dust surface area is dominated by the smallest grains, and in this case the dust flux enhancement due to grain trapping would be significantly greater than the mass enhancement factor $X$.

\subsection{Comparison to observations}
\label{subsec: comparisonToObservations}

\noindent We now compare predictions of the gas trap model to observational constraints on NIR excesses.

\subsubsection{Dust emission calculation}
\label{subsec: dustEmissionCalculation}

\noindent A major observational constraint on the nature of hot dust is that many systems have an excess NIR flux of around ${1 \; {\rm per \; cent}}$ of the stellar level, but the MIR flux is at least an order of magnitude lower and often undetected. In this section we model the flux arising from dust in the gas trapping scenario, in order to compare the model predictions to observational constraints.

We use \textsc{radmc} \citep{Dullemond2012} to model the combined flux arising from three populations of dust: trapped grains, dust migrating inwards from an external source, and unbound dust produced by collisions between trapped grains. We do not model a fourth population of grains transitioning from the migrating population to the trapped one, since this sublimation process occurs over very short timescales (Figure \ref{fig: evolutionWithContinuousGas}).

We model trapped dust as a population of equal-sized grains confined to a narrow disc of radius $r_{\rm s}$, width ${\Delta r / r_{\rm s} = 0.1}$ and half opening angle $5^\circ$, with number density at distance $r$ going as ${r^{-1}}$. We use a total trapped dust mass of ${10^{-11} \; {\rm M_\oplus}}$ for G2 stars and ${10^{-9} \; {\rm M_\oplus}}$ for A0 stars, as required to explain NIR observations \citep{Kirchschlager2017}. We only examine combinations of trapped dust sizes and locations that are consistent with the final grain parameters from our dynamical simulations (Figure \ref{fig: finalGrainParametersForDifferentGasDensities}).

The migrating dust population is assumed to form a continuous disc of varying density, stretching from the dust source down to the sublimation radius. For this analysis we assume the source to be located in the habitable zone (at 1 and ${6 \; {\rm au}}$ for G2 and A0 stars respectively), such that there is a significant MIR flux contribution. For G2 stars we include grains with radii from ${s_{\rm max} = 1 \; {\rm mm}}$ down to ${0.5 \; \mu {\rm m}}$ (just above the blowout size), in 11 logarithmically-spaced size bins. For A0 stars we include grain radii from ${1 \; {\rm mm}}$ down to ${5 \; \mu {\rm m}}$ across 6 logarithmically-spaced size bins. The migrating dust follows a size distribution with number density ${n(s) \propto s^{-p}}$, where we assume ${p = 3.5}$ (although the flux is independent of this value, as shown below). The mass inflow rate for each grain size is calculated as

\begin{equation}
\dot{M}(s) = \dot{M} \frac{\int_{s_{1}}^{s_{2}} m(s) n(s) {\rm d}s}{\int^{s_{\rm max}}_{s_{\rm bl}} m(s) n(s) {\rm d}s},
\label{eq: massInflowInEachGrainSize}
\end{equation}

\noindent where $m(s)$ is the mass of a grain of radius $s$, and the size bin has edges ${s_1 < s < s_2}$. The total mass input rate $\dot{M}$ is that required to maintain the given dust and gas population (Section \ref{subsec: dustInflowRate}). The region from the sublimation radius to the dust source is divided into annuli, and the total mass in each grain size across each annulus is calculated from Equations \ref{eq: prMigrationTime}, \ref{eq: inflowingMassInAnnulus} and \ref{eq: massInflowInEachGrainSize}. Note that the flux from a single-sized dust population of radius $s$ and total mass $M(s)$ scales roughly as ${M(s) / s}$, and Equations \ref{eq: prMigrationTime} and \ref{eq: inflowingMassInAnnulus} show that ${M(s) \propto 1 / \beta(s)}$, where ${1 / \beta(s) \propto s}$ for carbon grains above the blowout size. This means that the flux from migrating grains is roughly constant across all grain sizes, and so the results are largely independent of the chosen size distribution index $p$ and maximum grain size $s_{\rm max}$. We omit collisional processing of migrating material and so the fluxes from our migrating grains are lower limits, since some mass may be lost as unbound grains are created via collisions in migrating dust (e.g. \citealt{Rigley2020}).

Finally, the unbound dust is modelled assuming collisional remnants are created at the trap location, and are then blown out of the system. There are two unbound populations. Firstly, there are the larger remnants of originally trapped grains that have now eroded down to the blowout size. Since this erosion occurs over a grain's collisional lifetime $\tau_{\rm col}$, the total rate at which mass escapes in these larger unbound grains is ${\dot{M} = M_{\rm d} / \tau_{\rm col} (s_{\rm bl} / s_0)^3}$. We model this population assuming all such grains are at the blowout size. Secondly, there are the much smaller ejecta that are released through collisions; since these account for the remaining collisional mass loss, the rate at which mass escapes in these smaller unbound grains is ${\dot{M} = M_{\rm d} / \tau_{\rm col} \left[1-(s_{\rm bl} / s_0)^3\right]}$. We assume that these small ejecta follow a size distribution ${n(s) \propto s^{-3.5}}$, with a minimum grain radius of ${1 \; {\rm nm}}$. The maximum ejecta grain radius ${s_{\rm ej, max}}$ (the size of the second-largest fragment following the collision) is approximated such that for each collision there are no ejecta larger than ${s_{\rm ej, max}}$, i.e. ${\int_{s_{\rm ej, max}}^{\infty} n(s) {\rm d} s < 1}$; whilst this value should strictly be found from collisional simulations or impact experiments, this approximation should be sufficient for our purposes. We consider unbound grain radii in logarithmically-spaced size bins from ${1 \; {\rm nm}}$ to ${s_{\rm ej, max}}$, using 8 bins for G2 stars and 13 for A0 stars. As for the migrating grain population, we divide the star system into annuli and calculate the quantity of unbound grains of a given grain size in each annulus as ${M(s) = \dot{M}(s) \Delta t (s)}$. We use the $\dot{M}$ values calculated above, and take ${\Delta t(s)}$ as the time the escaping grain spends in each annulus. These times are interpolated directly from dynamical simulations of the unbound grains, similar to those shown on Figure \ref{fig: unboundGrainTrajectories}. The unbound grains are therefore subject to sublimation, radiation forces and gas drag. We truncate the unbound grain populations at ${2 \; {\rm au}}$ for G2 stars and ${10 \; {\rm au}}$ for A0 stars; increasing these distances would cause the escaping grain emission to peak farther into the MIR. Note that smaller ejecta released at the trapping radius rapidly sublimate before escaping, so provide a negligible flux contribution; the smallest ejecta that escape rather than sublimate have radii ${s \sim 0.1 \; \mu {\rm m}}$ for both star types.

We use the same Kurucz stellar spectra as in the earlier dynamical simulations, and the dust temperatures are interpolated from our pre-calculated grids as described in Section \ref{subsec: dustGrainProperties}. The dust opacities are calculated using the Bohren \& Huffman Mie code supplied with \textsc{radmc}, with the same optical constants for ${1000\degr{\rm C}}$ carbon as used in our dynamical simulations \citep{Jager1998}. We include thermal emission and simple isotropic scattering, although since the dominant emission is thermal, different scattering prescriptions do not significantly affect the results. We calculate all SEDs twice: once with the system face-on to the observer, and once with it edge-on.

\subsubsection{Dust emission results}
\label{subsec: dustEmissionResults}

\begin{figure*}
  \centering
   \includegraphics[width=16cm]{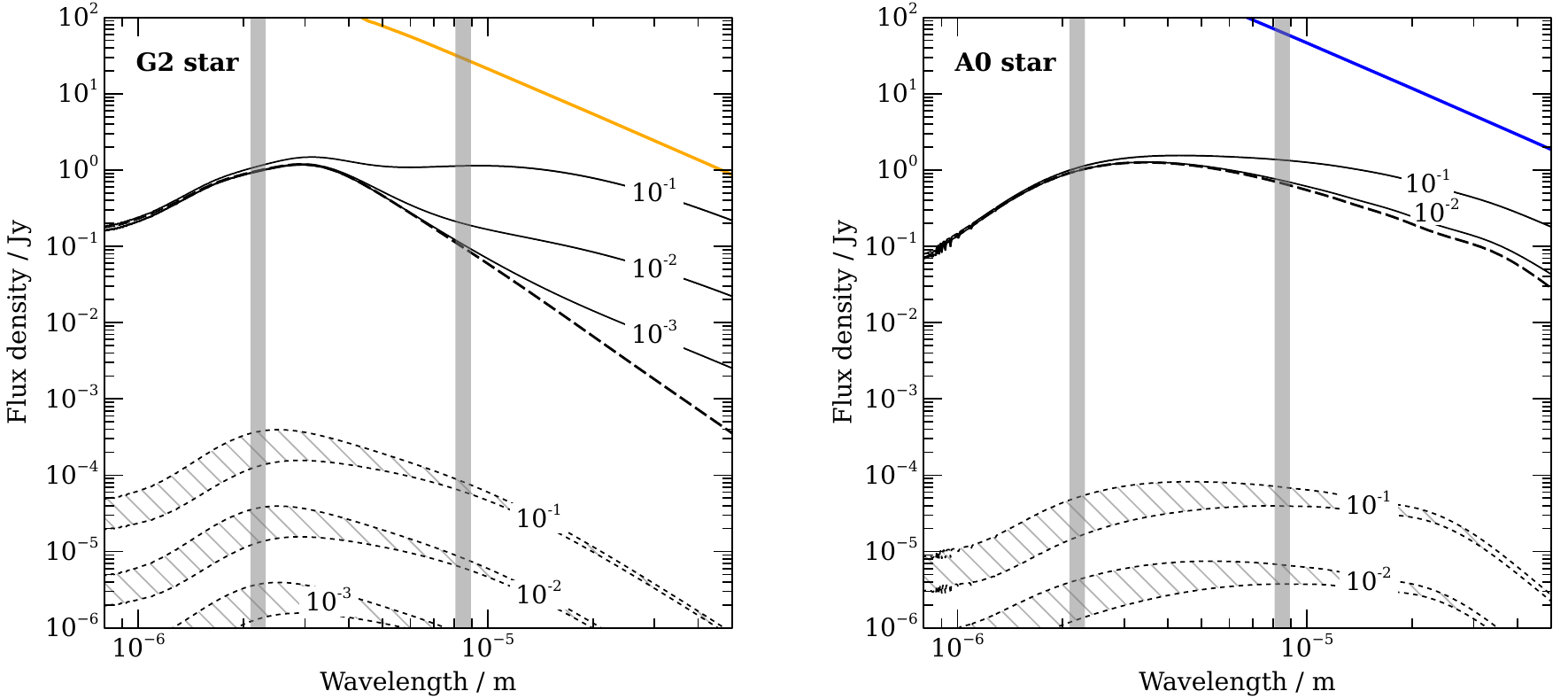}
   \caption{Simulated SEDs for the gas-trap scenario. Three dust populations are modelled: trapped grains, dust migrating inwards from an external source, and unbound collisional ejecta released by trapped grains. Solid black lines are combined fluxes for all three populations, for different values of the gas viscosity parameter $\alpha$. For high viscosities (${\alpha \sim 0.1}$) trapped grains have short lifetimes, since collisions rapidly deplete trapped dust; in this case large dust inflows are required, and the dust SED is dominated by migrating grains. For smaller viscosities the required dust inflow is reduced, and the dust SED tends towards trapped grains alone (dashed black lines). Escaping grains make negligible contributions (black dotted lines). Hatched regions show differences between edge-on and face-on systems. The G2 and A0 star fluxes are the thick yellow and blue lines, respectively. Vertical bands mark 2.2 and ${8.5 \; \mu {\rm m}}$; to reproduce typical observations, the ${2.2 \; \mu {\rm m}}$ dust flux should be at least ten times that at ${8.5 \; \mu {\rm m}}$. The model reproduces observations for Sun-like stars, provided ${\alpha \lesssim 10^{-2}}$. For A-type stars the ${8.5 \; \mu {\rm m}}$ flux is too large, because gas cannot trap sub-blowout grains. The ejecta fluxes qualitatively differ, since G2 ejecta sublimate before escaping, whilst larger A0 ejecta survive and are blown out of the system. The data are for gas densities at the sublimation radius of ${\rho_{\rm gas}(r_{\rm s}) = 10^{-19} \; {\rm g \; cm^{-3}}}$. The G2 star model shows ${10^{-11} \; {\rm M}_\oplus}$ of ${0.5 \; \mu {\rm m}}$ radius grains trapped at ${0.03 \; {\rm au}}$, with a dust source at ${1 \; {\rm au}}$. The A0 star model shows ${10^{-9} \; {\rm M}_\oplus}$ of ${5 \; \mu {\rm m}}$ radius grains trapped at ${0.2 \; {\rm au}}$, with a dust source at ${6 \; {\rm au}}$. All fluxes are scaled such that the ${2.2 \; \mu {\rm m}}$ flux of trapped grains alone is ${1 \; {\rm Jy}}$.}
   \label{fig: g2AndA0Spectra}
\end{figure*}

\noindent Figure \ref{fig: g2AndA0Spectra} shows typical spectra arising from the gas trap model. For a given trapped grain population, both the quantity of unbound grains and the mass inflow rate required to sustain the trapped population depend on the gas viscosity (Equations \ref{eq: massLossRateG2Star} and \ref{eq: massLossRateA0Star}); we therefore show SEDs for several $\alpha$ values. Higher gas viscosities lead to more violent collisions amongst trapped grains, and so higher mass inflow rates are required to maintain the trapped dust. In these cases migrating dust makes a significant contribution to the SED. For lower viscosities (${\alpha \lesssim 10^{-2}}$) the dust SED is dominated by emission from trapped grains. In all cases the flux from escaping unbound grains is negligible compared to that from trapped and migrating grains, since this dust either rapidly sublimates or escapes from the system.

We now use these simulated SEDs to examine how well the gas trap scenario reproduces observations. We compare the simulated fluxes at ${2.2 \; \mu {\rm m}}$ to those at ${8.5 \; \mu {\rm m}}$; these wavelengths are marked by grey bands on Figure \ref{fig: g2AndA0Spectra}, and are those at which many systems with NIR excesses have been observed (e.g. \citealt{Absil2013, Mennesson2014}). In order to reproduce typical observations, the ${2.2 \; \mu {\rm m}}$ NIR dust flux should be at least an order of magnitude larger than the ${8.5 \; \mu {\rm m}}$ MIR dust flux (e.g. \citealt{Kirchschlager2017}). 

For Sun-like stars we find that, provided gas turbulence is not too high, the gas trap mechanism can reproduce observations of NIR and MIR excesses. If ${\alpha \lesssim 10^{-2}}$, then Figure \ref{fig: g2AndA0Spectra} shows that the resulting NIR flux (predominantly from trapped sub-micron grains) outweighs the MIR flux (predominantly from grains migrating inwards from an external source) by an order of magnitude. Similarly, if the dust source is closer to the star (e.g. dust is deposited in the inner regions by comets, and migrates inwards from there) then the MIR flux is reduced and the gas trap model could reproduce observations even if ${\alpha \sim 1}$. 

These results agree with \citet{Kirchschlager2017}, who aimed to reproduce NIR and MIR excesses assuming pure graphite dust populations of a single size and distance. They were able to constrain hot dust locations for two Sun-like stars: ${0.013 - 0.06 \: {\rm au}}$ for hot dust around ${\tau \; \rm Ceti}$ (${0.46 \; {\rm L}_\odot}$), and ${0.035 - 0.52 \: {\rm au}}$ for that around ${{\rm HD} \; 22484}$ (${3.0 \; {\rm L}_\odot}$). Since the gas trap location is set by dust temperature, ${r_{\rm s} \propto \sqrt{L_*}}$ and so these inferred dust locations are consistent with our prediction that gas trapping occurs at distances of ${0.02 - 0.03 \: {\rm au}}$ from G2 stars. If the observed dust around ${\tau \; {\rm Ceti}}$ is at the outermost location expected from gas trapping, then the hot grains must be ${\lesssim 0.5 \: \mu {\rm m}}$ in radius (their Figure 10), in agreement with our G2 star results. If it is located closer in then \citet{Kirchschlager2017} cannot constrain the grain size. The grain size is also unconstrained for ${{\rm HD} \; 22484}$. Our predictions for dust trapped in gas can therefore replicate the observational results of \citet{Kirchschlager2017} for Sun-like stars.

However, for A0 stars we find that the gas trap as modelled here is unable to produce a NIR flux that is large enough relative to MIR to explain observations. Even for low $\alpha$ values, the MIR flux from the trapped population alone is at least ${60 \; {\rm per cent}}$ of the NIR flux. The reason for this discrepancy is that gas cannot trap grains smaller than the blowout size, which for A0 stars is slightly smaller than ${5 \; \mu {\rm m}}$. However, since dust also cannot be trapped interior to the sublimation region, the trapped grains cannot get hot enough to explain the high NIR flux relative to that at MIR. Whilst smaller grains emitting primarily in the NIR would be created through collisions at the trap location, their flux contribution is small since they either rapidly sublimate or quickly blow out of the system. 

Again, our results agree with \citet{Kirchschlager2017}. Of the 9 A stars they analysed, all have inferred hot dust locations compatible with our result of ${0.1 - 0.2 \: {\rm au}}$ for an A0 star (again, having scaled the trap distance as ${r_{\rm s} \propto \sqrt{L_*}}$). The gas trap model can therefore reproduce the inferred location of hot dust around A0 stars. However, \citet{Kirchschlager2017} infer that the observed hot dust is smaller than ${0.1 - 1 \: \mu {\rm m}}$ (their Figure 7), whilst we find that gas cannot trap grains smaller than ${5 \: \mu {\rm m}}$ around an A0 star. This discrepancy regarding the sizes of grains trapped around A stars is discussed in Section \ref{sec: discussion}.

Finally, in addition to grain size and location, several other observational constraints can be compared to predictions from the gas trap model. First, \citet{Kirchschlager2017} show that hot dust location scales with stellar temperature; this is naturally reproduced by gas trapping because grains are trapped around the sublimation distance, which increases with stellar luminosity. Second, observations suggest that hot dust populations have steep size and density distributions, with an overabundance of small grains located in a narrow region close to the star \citep{diFolco2007, Defrere2011, Lebreton2013}. The gas trap mechanism reproduces both constraints. A steep size distribution is predicted from gas trapping because trapped grains are close to the blowout size, and larger grains are absent since they sublimate down to this size. Likewise since grains of all initial sizes drift to the same radial distance when they get trapped (Figure \ref{fig: convergingGrainParametersForSingleGasDensity}), the trapped population would occupy a narrow distance range. Finally, no correlation is observed between detected hot excesses and warm or cold excesses around stars \citep{MillanGabet2011, Ertel2014, Mennesson2014, Ertel2018, Ertel2020}. Since gas trapping requires low dust inflow rates, only small quantities of source material are required outside the sublimation region. 

In summary, provided that gas turbulence is not too high or that dust is supplied interior to the habitable zone by comets, gas trapping can produce the hot dust inferred around Sun-like stars. For A stars the dust location is reproduced, but trapped grains are ${5-10}$ times larger than those inferred from observations. This is discussed in Section \ref{sec: discussion}. Aside from this, the gas trap mechanism is compatible with several other observational results.

\subsubsection{Observability of the gas}
\label{subsec: gasEmission}

\noindent The model assumes atomic carbon gas is present in sufficient quantities to trap dust, and here we investigate whether such gas is observable in emission. The relevant emission lines are CI (610 and ${370 \: \mu {\rm m}}$) and CII (${158 \: \mu {\rm m}}$), falling within the ranges of ALMA and SOFIA, respectively. For an order of magnitude estimate we use a similar method to \citet{Roberge2013}, assuming gas is in local thermal equilibrium (LTE) and optically thin.

Gas mass is related to the flux of an emission line by

\begin{equation}
M_{\rm gas} = \frac{4 \pi d^2 \mu m_{\rm u} \mathcal{F}_{{\rm U,L}}}{h \nu_{{\rm U, L}} A_{{\rm U,L}} x_{\rm U}},
\label{eq: gasMassFromEmissionLineFlux}
\end{equation}

\noindent where $d$ is the distance from Earth, $h$ is Planck's constant, ${\mathcal{F}_{{\rm U,L}}}$, ${\nu_{{\rm U,L}}}$ and ${A_{{\rm U, L}}}$ are the flux, rest frequency and Einstein A coefficient, respectively, associated with the transition from upper energy level $\rm{U}$ to lower level $\rm{L}$, and $x_{\rm U}$ is the fraction of molecules in the upper level. In LTE, $x_{\rm U}$ is

\begin{equation}
x_{\rm U} = \frac{2 J_{\rm U} + 1}{Q(T_{\rm gas})} \exp\left(- \frac{E_{\rm U,L}}{k_{\rm B} T_{\rm gas}} \right),
\label{eq: fractionOfMoleculesInUpperEnergyLevel}
\end{equation}

\noindent where $J_{\rm U}$ is the angular momentum quantum number of the upper level, $E_{\rm U,L}$ is the energy difference between the upper and lower levels, and $Q(T_{\rm gas})$ is the partition function:

\begin{equation}
Q(T_{\rm gas}) = g_L + g_{\rm U} \exp\left(- \frac{E_{\rm U,L}}{k_{\rm B} T_{\rm gas}} \right),
\label{eq: partitionFunction}
\end{equation}

\noindent where $g_{\rm i}$ is the statistical weight of the ${i{\rm th}}$ energy level (e.g. \citealt{RiviereMarichalar2014}). 

We now estimate the minimum gas mass observable by ALMA, using the online sensitivity calculator\footnote{\url{https://almascience.eso.org/proposing/sensitivity-calculator}}. We consider the low spectral resolution Time Division Mode (TDM), to maximise gas detection potential. In TDM the spectral resolution is ${31.2 \: {\rm MHz}}$ (Table A-4 in the proposer's guide\footnote{\url{https://almascience.eso.org/documents-and-tools/cycle8/alma-proposers-guide}}). We assume that 40 antennas are used in dual polarization mode to observe a source at declination $-35^\circ$, for a total integration time of ${1 \: {\rm hour}}$, with the automatic choice for water vapour column density; similar parameters are used in \citet{Kral2017Gas}. The resulting $5\sigma$ ALMA sensitivities are ${3 \times 10^{-21}}$ and ${2 \times 10^{-20} \: {\rm W m}^{-2}}$ at ${610}$ and ${370 \: \mu {\rm m}}$, respectively. Assuming ${1850 \: {\rm K}}$ gas and using atomic data from Table \ref{tab: atomicData} \citep{Schoier2005}, Equation \ref{eq: gasMassFromEmissionLineFlux} shows that the minimum CI gas masses at ${10 \: {\rm pc}}$ detectable by ALMA are ${7 \times 10^{-7}}$ and ${8 \times 10^{-7} M_\oplus}$ at 610 and ${370 \: \mu {\rm m}}$, respectively.

For the CII ${158 \: \mu{\rm m}}$ emission line, we estimate the minimum gas mass observable using FIFI-LS on SOFIA. From Figure 3-4 in the observer's handbook\footnote{\sofiaFIFIHandbook}, a ${15 \: {\rm minute}}$ observation results in a $4\sigma$ detectable line flux of ${2 \times 10^{-17} \: {\rm W m}^{-2}}$ at ${158 \: \mu{\rm m}}$. Since sensitivity is proportional to $t_{\rm int}^{-0.5}$, where $t_{\rm int}$ is integration time, the $5\sigma$ sensitivity for a ${1 \: {\rm hour}}$ integration is ${1 \times 10^{-17} \: {\rm W m}^{-2}}$. Equation \ref{eq: gasMassFromEmissionLineFlux} therefore yields the minimum CII gas mass at ${10 \: {\rm pc}}$ detectable by SOFIA as ${3 \times 10^{-5} M_\oplus}$.

\begin{table*}
\begin{tabular}{c c c c c c c c}
\hline
Ionisation state & Transition & Wavelength / $\mu$m & $A_{{\rm U, L}} \: / \: {\rm s}^{-1}$ & ${E_{\rm U,L}/k_{\rm B} \: / \: {\rm K}}$ & $g_{\rm U}$ & $g_L$ \\
\hline

CI & ${^3P_{1} \rightarrow {^3}P_{0}}$ & 610 & ${7.9 \times 10^{-8}}$ & 23.62 & 3 & 1 \\

CI & ${^3P_{2} \rightarrow {^3}P_{1}}$ & 370 & ${2.7 \times 10^{-7}}$ & 62.46 & 5 & 3 \\

CII & ${^2P_{3/2} \rightarrow {^2}P_{1/2}}$ & 158 & ${2.3 \times 10^{-6}}$ & 91.21 & 4 & 2 \\

\hline
\end{tabular}
\caption{Atomic data for various carbon transitions, from \citet{Schoier2005}. Data are used to calculate gas emission in Section \ref{subsec: gasEmission}.}
\label{tab: atomicData}
\end{table*}

The minimum carbon gas masses detectable by ALMA and SOFIA ($10^{-7}$ to ${10^{-5} \: M_\oplus}$) are much larger than those required to trap dust, which could be as low as ${10^{-12} \: M_\oplus}$ (Figure \ref{fig: trappingAndGrowthInContinuousGasDensity}). We therefore conclude that gas trapping may operate with sub-detection levels of gas, and that non-detections of gas emission around stars with NIR excesses would not invalidate the  model. 

\citet{Rebollido2018, Rebollido2020} detected hot CaII and NaI gas in absorption around some stars with NIR excesses. The absorption is variable, and is probably produced close to stars by Falling Evaporating Bodies (FEBs). For such gas to be detected implies that it exists in greater quantities than our paper considers. We argue that this observed gas is a separate entity, potentially coexisting with a much smaller quantity of dust-trapping gas, but not necessarily associated with it. This is because the large variability implies that the gas is not contained in a stable disc capable of trapping grains for long periods, and high variability would not be expected from a constant inflow of sublimating dust. Indeed, FEB simulations show that released CaII gas does not form a disc capable of maintaining trapped dust, but is quickly blown away by stellar radiation \citep{Beust1990}. Whilst it may be possible to trap hot dust in the gas released by comets, it is unclear whether the gas would form a stable disc that could trap dust for long periods. We therefore argue that variable hot gas detections of \citet{Rebollido2018, Rebollido2020} are associated with FEBs, and that if the trapping mechanism occurs in those systems, it operates with distinct, undetected gas.

The gas trap model is not incompatible with FEBs, and could benefit from them. Dust released by FEBs near the star takes less time to reach the sublimation region than dust from the habitable zone, so trapped grains could have shorter lifetimes without a significantly decreased NIR-to-MIR flux ratio. \citet{Rebollido2020} suggest a tentative trend for variable gas detections in systems with NIR excesses, so FEBs could help supply dust to the inner system. Additionally, whilst the large and variable quantities of gas released by comets may struggle to trap bound grains, they may significantly slow the escape of unbound grains released close to the star (Section \ref{subsec: behaviourOfSubblowoutGrains}).

\subsection{Validity of gas model}
\label{subsec: validityOfgasModel}

\noindent Given the very small gas quantities investigated in this paper, we must check that such tenuous gas obeys our hydrodynamic prescription. Following \citet{Marino2020}, hydrodynamic treatment is valid if the mean free path of gas molecules, $\lambda$, is much smaller than the disc scale height $H$. Combining Equations \ref{eq: scaleHeight} and \ref{eq: meanFreePath}, ${\lambda \ll H}$ if

\begin{equation}
\rho_{\rm gas} \gg \frac{1}{\sigma}\left(\frac{\mu m_{\rm u}}{r}\right)^{3/2}\left({\frac{G M_*}{2 k_{\rm B} T_{gas}}}\right)^{1/2}.
\label{eq: gasDensityForHydrostaticTreatmentValid}
\end{equation}

\noindent Substituting typical values for our discs and ${\sigma \sim 10^{-16} \: {\rm cm}^2}$ for atomic carbon yields that the mean free path is only much smaller than the disc scale height if ${\rho_{\rm gas} \gg 10^{-17} \: {\rm g \: cm}^{-3}}$. This is a typical gas density required to trap dust, hence gas behaviour may be more complex than we assumed. However, gas close to the star would have a non-negligible ionisation fraction, in which case \citet{Marino2020} argue that the mean free path would be orders of magnitude shorter owing to larger ionised cross sections. We therefore argue that our gas model is valid for the densities considered, but possibly at the edge of the hydrodynamic limit. Applying our results to more tenuous gas should be done with caution.
\section{Discussion}
\label{sec: discussion}

\noindent We have investigated dust trapping by gas as a potential mechanism to explain NIR excesses around main-sequence stars. We showed that gas released by sublimation can effectively trap dust close to the star, and that only small quantities of gas are required. The trapped dust would have a steep size distribution and exist around the sublimation region, in agreement with inferences from observations. We showed that gas trapping of solid carbon grains can reproduce the hot dust sizes and locations inferred from observations of Sun-like stars. For A-type stars the predicted location of hot dust also matches that inferred from observations; however, in this case our modelled grains are an order of magnitude too large. This means that, whilst we were able to produce dust emission at ${2.2 \; \mu {\rm m}}$ that is twice that at ${8.5 \; \mu {\rm m}}$, we were unable to increase this factor to ten times as required by observations. The issue arises because gas cannot trap grains smaller than the blowout size, and so the ${\gtrsim 5 \; \mu {\rm m}}$ grains trapped around A stars are not hot enough to emit strongly in the NIR without also exceeding the observationally-allowed MIR flux.

This grain size discrepancy for A stars poses a difficulty for the gas trap model. Whilst we showed that grains of the required smaller sizes would be produced through collisions between trapped dust, these unbound grains are too short-lived to significantly contribute to the overall dust emission; they either rapidly sublimate or are blown out of the system. Gas drag could slow the escape of ejecta, but the smallest grains (those slowed most by gas) sublimate long before they would escape the system, and so this mechanism does not significantly increase the emission from sub-blowout grains.

It is possible that the results could change for different grain properties, such as composition or porosity, or that different sublimation prescriptions could allow hotter trapped grains to survive closer to the star. It is also possible that multiple gas traps exist at various distances from the star, since different grain compositions would sublimate at different radii. However, we consider these unlikely solutions. Highly-porous grains have larger blowout sizes and lower temperatures \citep{Kirchschlager2013, Brunngraber2017}, which is the opposite of what is required. Similarly, a blowout-sized grain emitting as a blackbody would need to be considerably hotter than those trapped around A stars in our simulations to reproduce observations, and it is difficult to explain how such dust could survive sublimation for a significant length of time, regardless of composition or sublimation prescription.

\begin{figure}
  \centering
   \includegraphics[width=7cm]{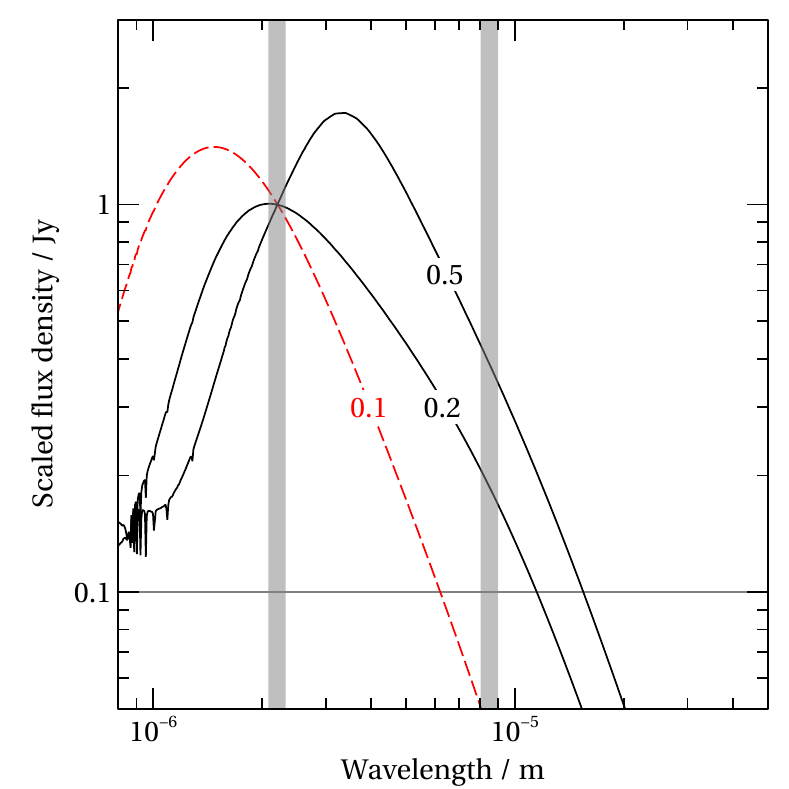}
   \caption{Regardless of origin, unbound grains released near A0 stars struggle to reproduce observations. The plot shows fluxes from unbound carbon grains of a single size released at the sublimation radius around an A0 star, with no gas present. Labels are grain radii in microns. Grey bands show 2.2 and ${8.5 \; \mu {\rm m}}$. Fluxes are scaled such that the ${2.2 \; \mu {\rm m}}$ flux is ${1 \; {\rm Jy}}$, so to reproduce observations the ${8.5 \; \mu {\rm m}}$ flux should be below ${0.1 \; {\rm Jy}}$ (horizontal line). Black solid lines show grains that survive sublimation and blow out of the system, which do not reproduce observations. The red dashed line shows grains that sublimate before being blown out, which reproduce observations but sublimate just minutes after creation. The plot only includes emission from grains within ${10 \; {\rm au}}$; increasing this cutoff causes the black SEDs to peak at longer wavelengths, whilst the red SED is unaffected.}
   \label{fig: subBlowoutGrains}
\end{figure}

The fundamental problem is that it is unclear how to get dust around A stars to be either small enough or hot enough for their NIR emission to be an order of magnitude larger than their MIR emission, without the grains either rapidly sublimating or being blown out of the system. It is important to note that this problem plagues \textit{all} current explanations of NIR excesses. Even in the cometary delivery model (where star-grazing comets release dust at very small distances, thus eliminating flux from inwardly migrating grains), the resulting MIR flux is too large relative to that at NIR for A-type stars (e.g. \citealt{Sezestre2019}). Regardless of production mechanism, for ${2.2 \; \mu {\rm m}}$ dust emission around an A0 star to be an order of magnitude larger than that at ${8.5 \; \mu {\rm m}}$ requires one of the following possibilities:

\begin{itemize}
\item Observed dust is smaller than the blowout size and unbound (i.e. ${s < 0.5 \; \mu {\rm m}}$). However, such grains rapidly evolve through radiation pressure or sublimation. Unbound grains large enough to survive sublimation are unlikely to be the solution, since these rapidly cool as they blow out of the system and contribute significantly to MIR flux, despite their small size. We show this on Figure \ref{fig: subBlowoutGrains}, where sub-blowout grains released close to an A0 star do not produce NIR excesses an order of magnitude larger than those at MIR if they survive sublimation. Such grains could only be a solution if their distribution is somehow truncated at several au. Smaller grains that rapidly sublimate do not suffer from this problem;  Figure \ref{fig: subBlowoutGrains} shows that such grains can produce NIR fluxes an order of magnitude larger than those at MIR, because the grains are concentrated at small distances. However, such grains sublimate within a matter of minutes, so would require huge supply rates in order to maintain ${\sim 10^{-9} \; {\rm M_\oplus}}$ populations close to an A0 star. It is difficult to see how any production mechanism could maintain adequate populations of such grains.

\item Observed dust is larger than the blowout size and does not escape, but is near the Rayleigh-Jeans limit at the observed wavelengths. Since the wavelength $\lambda_{\rm peak}$ at which the emission per unit frequency is largest goes as ${{\lambda_{\rm peak} = 5100 \; \mu {\rm m} / (T / {1 \; \rm K})}}$, this implies that dust temperatures are ${\gtrsim 2300 \; {\rm K}}$; this would require blowout-sized carbon grains to be located at ${0.08 \; {\rm au}}$ from an A star, which is roughly half the sublimation distance that arises from our models. At this distance, Equations \ref{eq: sublimationRateNoGas} and \ref{eq: gasDensityAtSaturationPressure} show that such grains would fully sublimate within just ${\sim 10 \; {\rm hours}}$, and so a stable hot dust population of ${10^{-9} \; {\rm M_\oplus}}$ around an A star would require 1000 comets of radius ${1 \; \rm km}$ to fully sublimate interior to the sublimation region \textit{per day}. Again, it is difficult to see how such a large inflow could be sustained.

\item Observed dust has some composition with a strong spectral feature in the NIR, or that is better able to withstand sublimation. However, many dust compositions have already been explored in the literature (e.g. \citealt{Kirchschlager2017}), and carbonaceous grains (with no strong NIR features) still appear to be the most likely composition.

\item Some additional, as yet untested mechanism exists that allows very hot dust to survive for sufficiently long times close to A0 stars, or reduces the relative emission from escaping grains.

\end{itemize}

The above considerations mean that no mechanism has so far been able to maintain the required quantities of hot dust close to an A0 star to be able to reproduce observations. Trapping models, such as that considered in this paper, would have to somehow prevent blowout-sized grains from escaping, whilst simultaneously protecting them from sublimation. Likewise, supply models where carbonaceous dust is created close to the star and blows out of the system (such as cometary delivery) cannot produce NIR fluxes that are larger than those at MIR, owing to the non-negligible flux arising from escaping grains that have cooled. It may be that a combination of trapping and supply mechanisms are required to explain NIR excesses.

\vspace{2cm}

\section{Conclusions}
\label{sec: conclusions}

\noindent Since 2006 some main-sequence stars have been known to host excess near-infrared emission. Surveys detect this phenomenon for one fifth of stars, across various spectral types and ages. The excesses are commonly interpreted as populations of small, hot dust grains very close to the stars. However, the presence of such hot grains in copious amounts is a mystery, since they should rapidly sublimate and blow out of the system. Many potential scenarios have been explored in the literature, but to date none satisfactorily explain both the phenomenon and its ubiquity. 

We investigate a simple mechanism to generate excesses: dust migrating inwards under radiation forces sublimates near the star, releasing gas which traps subsequent grains. The mechanism requires neither specialised system architectures nor high dust supply rates, and could operate across diverse stellar types and ages. It can significantly enhance dust masses close to the star, reducing the mass inflow rates required to generate infrared excesses.

Although we have been unable to fully reproduce all observations using the gas trap model, we have shown that the mechanism can naturally reproduce observed excesses around Sun-like stars, and that the trapped dust location scales with star luminosity with grains following a steep size distribution, as inferred from observations. For A0 stars we have not been able to produce dust emission at ${2.2 \; \mu {\rm m}}$ that is ten times that at ${8.5 \; \mu {\rm m}}$ as required by observations, but we have shown that trapped grains in low-viscosity gas can produce a ${2.2 \; \mu {\rm m}}$ flux that is twice that at ${8.5 \; \mu {\rm m}}$. Our simulated trapped grains around A0 stars are ${5-10}$ times larger than those required to reproduce observations.

Finally, we have identified a number of significant problems that any hot dust explanation of NIR excesses must overcome. Further progress with any hot dust explanation for A-type stars requires a means for grains to become very hot without either rapidly sublimating or being blown out of the system.

\section*{Acknowledgements}
\noindent We thank F. Kirchschlager and T. L\"ohne for discussions, as well as the anonymous referee whose comments substantially improved the paper. This research was supported by the \textit{Deutsche Forschungsgemeinschaft} (DFG), grants Kr 2164/14-2 and Kr 2164/15-2, in the framework of the DFG
Research Unit FOR 2285 ``Dust in Planetary Systems''.

\section*{Data availability}
\noindent The data underlying this article will be shared upon reasonable request to the corresponding author.


\bibliographystyle{mn2e}
\bibliography{bib_hotDustGasTrapping}


\label{lastpage}

\end{document}